\documentclass[superscriptaddress,prc,aps,twocolumn,showpacs]{revtex4-1}
\usepackage{amsmath,amssymb}
\usepackage{graphicx}
\usepackage{hyperref}
\usepackage{cleveref}
\usepackage{booktabs}
\usepackage{color}
\usepackage{mathtools}
\usepackage{amsmath}

\DeclareMathOperator{\CL}{CL}

\newcommand{\Mainz}[1]
{\affiliation{Institut f\"ur Kernphysik, Johannes Gutenberg-Universit\"at Mainz, D-55099 
Mainz, Germany}}

\newcommand{\Bonn}[1]
{\affiliation{Helmholtz-Institut f\"ur Strahlen- und Kernphysik, Universit\"at Bonn, D-53115 
Bonn, Germany}}

\newcommand{\Kent}[1]
{\affiliation{Kent State University, Kent, Ohio 44242-0001, USA}}

\newcommand{\Glasgow}[1]
{\affiliation{SUPA School of Physics and Astronomy, University of Glasgow, Glasgow G12 8QQ, 
United Kingdom}}

\newcommand{\Dubna}[1]
{\affiliation{Joint Institute for Nuclear Research, 141980 Dubna, Russia}}

\newcommand{\Pavia}[1]
{\affiliation{INFN Sezione di Pavia, I-27100 Pavia, Italy}}

\newcommand{\GWU}[1]
{\affiliation{The George Washington University, Washington, DC 20052-0001, USA}}

\newcommand{\Dalhousie}[1]
{\affiliation{Dalhousie University, Halifax, Nova Scotia B3H 4R2, Canada}}

\newcommand{\Halifax}[1]
{\affiliation{Department of Astronomy and Physics, Saint Mary's University, Halifax, Nova 
Scotia B3H 3C3, Canada}}

\newcommand{\UniPavia}[1]
{\affiliation{Dipartimento di Fisica, Universit\`a di Pavia, I-27100 Pavia, Italy}}

\newcommand{\Basel}[1]
{\affiliation{Departement f\"ur Physik, Universit\"at Basel, CH-4056 Basel, Switzerland}}

\newcommand{\Edinburgh}[1]
{\affiliation{SUPA School of Physics, University of Edinburgh, Edinburgh EH9 3JZ, United 
Kingdom}}

\newcommand{\INR}[1]
{\affiliation{Institute for Nuclear Research, 125047 Moscow, Russia}}

\newcommand{\Sackville}[1]
{\affiliation{Mount Allison University, Sackville, New Brunswick E4L 1E6, Canada}}

\newcommand{\Regina}[1]
{\affiliation{University of Regina, Regina, Saskatchewan S4S 0A2, Canada}}

\newcommand{\Zagreb}[1]
{\affiliation{Rudjer Boskovic Institute, HR-10000 Zagreb, Croatia}}

\newcommand{\Amherst}[1]
{\affiliation{University of Massachusetts, Amherst, Massachusetts 01003, USA}}

\newcommand{\UCLA}[1]
{\affiliation{University of California Los Angeles, Los Angeles, California 90095-1547, 
USA}}

 \newcommand{\Jerusalem}[1]
{\affiliation{Racah Institute of Physics, Hebrew University of Jerusalem, Jerusalem 91904, Israel}}

\begin{document}

\title{Measurement of the decay $\eta^{\prime}\to\pi^{0}\pi^{0}\eta$ at MAMI}

\author{P.~Adlarson}\thanks{corresponding author, e-mail: padlarso@uni-mainz.de}\Mainz \\
\author{F.~Afzal}\Bonn \\
\author{Z.~Ahmed}\Regina \\
\author{C.~S.~Akondi}\Kent \\
\author{J.~R.~M.~Annand}\Glasgow  \\
\author{H.~J.~Arends}\Mainz \\
\author{R.~Beck}\Bonn \\
\author{N.~Borisov}\Dubna \\
\author{A.~Braghieri}\Pavia \\
\author{W.~J.~Briscoe}\GWU \\
\author{F.~Cividini}\Mainz \\
\author{C.~Collicott}\Dalhousie \\ \Halifax \\
\author{S.~Costanza}\Pavia \\ \UniPavia \\
\author{A.~Denig}\Mainz \\
\author{E.~J.~Downie}\Mainz \\ \GWU \\
\author{M.~Dieterle}\Basel \\
\author{M.~I.~Ferretti Bondy}\Mainz \\
\author{S.~Gardner}\Glasgow \\
\author{S.~Garni}\Basel \\
\author{D.~I.~Glazier}\Glasgow \\ \Edinburgh \\
\author{D.~Glowa}\Edinburgh \\
\author{W.~Gradl}\Mainz \\
\author{G.~Gurevich}\INR \\
\author{D.~J.~Hamilton}\Glasgow \\
\author{D.~Hornidge}\Sackville \\
\author{G.~M.~Huber}\Regina \\
\author{A.~K\"aser}\Basel\\
\author{V.~L.~Kashevarov}\Mainz \\ \Dubna \\
\author{S.~Kay}\Edinburgh \\
\author{I.~Keshelashvili}\Basel\\
\author{R.~Kondratiev}\INR \\
\author{M.~Korolija}\Zagreb \\
\author{B.~Krusche}\Basel \\
\author{A.~Lazarev}\Dubna \\
\author{J.~Linturi}\Mainz \\
\author{K.~Livingston}\Glasgow \\
\author{I.~J.~D.~MacGregor}\Glasgow \\
\author{D.~M.~Manley}\Kent \\ 
\author{P.~P.~Martel}\Mainz \\ \Sackville  \\
\author{J.~C.~McGeorge}\Glasgow \\
\author{D.~G.~Middleton}\Mainz \\ \Sackville \\
\author{R.~Miskimen}\Amherst \\
\author{A.~Mushkarenkov}\Pavia \\ \Amherst \\
\author{A.~Neganov}\Dubna \\
\author{A.~Neiser}\Mainz \\ 
\author{M.~Oberle}\Basel \\
\author{M.~Ostrick}\Mainz \\  
\author{P.~Ott}\Mainz \\   
\author{P.~B.~Otte}\Mainz \\
\author{B.~Oussena}\Mainz \\ \GWU \\
\author{D.~Paudyal}\Regina \\
\author{P.~Pedroni}\Pavia \\
\author{A.~Polonski}\INR \\  
\author{S.~Prakhov}\Mainz \\ \UCLA \\
\author{A.~Rajabi}\Amherst \\
\author{G.~Ron}\Jerusalem \\
\author{T.~Rostomyan}\Basel \\
\author{A.~Sarty}\Halifax \\  
\author{C.~Sfienti}\Mainz \\
\author{V.~Sokhoyan}\Mainz \\ \GWU \\
\author{K.~Spieker}\Bonn \\
\author{O.~Steffen}\Mainz \\
\author{I.~I.~Strakovsky}\GWU \\
\author{T.~Strub}\Basel \\
\author{I.~Supek}\Zagreb \\
\author{A.~Thiel}\Bonn \\
\author{M.~Thiel}\Mainz \\
\author{L.~Tiator}\Mainz \\  
\author{A.~Thomas}\Mainz \\   
\author{M.~Unverzagt}\Mainz \\
\author{Y.~A.~Usov}\Dubna \\
\author{S.~Wagner}\Mainz \\
\author{D.~P.~Watts}\Edinburgh \\
\author{D.~Werthm\"uller}\Glasgow \\ \Basel \\
\author{J.~Wettig}\Mainz \\
\author{L.~Witthauer}\Basel \\ 
\author{M.~Wolfes}\Mainz \\
\author{R.~L.~Workman}\GWU \\
\author{L.~A.~Zana}\Edinburgh \\

\collaboration{A2 Collaboration at MAMI}

\date{\today}

\begin{abstract}
An experimental study of the $\eta'\to \pi^0\pi^0\eta \to 6\gamma$ decay
has been conducted with the best up-to-date statistical accuracy, by
measuring $\eta'$ mesons produced in the $\gamma p \to \eta' p$ reaction
with the A2 tagged-photon facility at the Mainz Microtron, MAMI.
The results obtained for the standard parametrization of the $\eta'\to \pi^0\pi^0\eta$
matrix element are consistent with the most recent results for $\eta'\to\pi\pi\eta$ decays,
 but have smaller uncertainties. The available statistics and experimental resolution allowed,
for the first time, an observation of a structure below the
$\pi^+\pi^-$ mass threshold, the magnitude and sign of which,
checked within the framework of the nonrelativistic effective-field theory,
demonstrated good agreement with the cusp that was predicted
based on the $\pi\pi$ scattering length combination, $a_0-a_2$,
extracted from $K \to 3\pi$ decays.
\end{abstract}


\maketitle

\section{Introduction}
\label{Intro}
The $\eta'$ meson and its decay modes play an important role in understanding Quantum ChromoDynamics (QCD) and related theoretical models, which allow a test of the pion scattering lengths via $\eta'\to\pi\pi\eta$ decays~\cite{Kubis:2009sb}. Although, in the isospin limit, both the $\eta'\to \pi^0\pi^0\eta$ and $\eta'\to \pi^+\pi^-\eta$ decay amplitudes are the same, they become different because of strong final-state interactions, which are also expected
to create a pronounced cusp in the neutral-decay spectrum at the $\pi^{+}\pi^{-}$ mass threshold. A similar cusp was first seen by NA48/2 in the $K^{+}\to \pi^+\pi^0\pi^0$ decay and then used to extract the S-wave $\pi\pi$ scattering length combination $a_{0} - a_{2} $~\cite{Batley:2005ax, Cabibbo:2004gq, Cabibbo:2005ez, Colangelo:2006va, Gamiz:2006km, Bissegger:2008ff}. The cusp was subsequently studied in $K_L\to \pi^0\pi^0\pi^0$~\cite{Cabibbo:2005ez, Gamiz:2006km, Bissegger:2007yq, Abouzaid:2008aa} and $\eta\to \pi^0\pi^0\pi^0$ decays~\cite{Ditsche:2008cq, Gullstrom:2008sy, Adolph:2008vn, Unverzagt:2008ny, Prakhov:2008ff}. The cusp predicted for $\eta'\to \pi^0\pi^0\eta$ ~\cite{Kubis:2009sb} has not been observed experimentally so far.

Another $\eta'\to \pi^0\pi^0\eta$ test of QCD could also be made by comparing it to the isospin-violating $\eta'\to3\pi^0$ decay, which gives access to the light quark masses and the mixing properties of $\pi^{0}$ and $\eta$ mesons~\cite{Escribano:2010wt, Bickert:2016fgy}. As an initial hypothesis~\cite{Gross:1979ur}, the amplitudes of $\eta'\to \pi^0\pi^0\pi^0$ and $\eta'\to \pi^0\pi^0\eta$ can be related as
\begin{equation}
M(\eta'\to \pi^0\pi^0\pi^0) = 3\epsilon \cdot M(\eta'\to \eta\pi^0\pi^0), 
\label{eqn:etapipito3pi}
\end{equation}
where $\epsilon = (\sqrt{3}/4)(m_d - m_u)/(m_s - \hat{m})$ is the $\pi^{0}$-$\eta$ mixing angle, $m_{q}$ is the mass of quark $q$, and $\hat{m}$ is the averaged mass of quarks $u$ and $d$. Equation~(\ref{eqn:etapipito3pi}) assumes
that the $\eta'\to \pi^0\pi^0\pi^0$ decay occurs entirely through $\pi^{0}$-$\eta$ mixing in the $\eta'\to \eta\pi^0\pi^0$ decay. According to Ref.~\cite{Borasoy:2006uv}, such an assumption is too strong, but there should still be a nonnegligible contribution from $\eta\pi \to \pi\pi$ rescattering, which can, e.g., be described through dispersion relations~\cite{Isken:2017dkw}.
 
From the $\eta'\to\pi\pi\eta$ decay, one could also learn about QCD-related models and, in particular, the low-energy effective-field theory of QCD, Chiral Perturbation Theory (ChPT)~\cite{Gasser:1984gg}. In ChPT, the pseudoscalar singlet $\eta_{1}$, associated with the physical state $\eta'$, is not included explicitly (but implicitly through low-energy constants) due to the U(1)$_{A}$ anomaly, which also renders it massive in the chiral limit of massless quarks. To include the singlet, ChPT is extended by going to a large number of color charges (large-$N_{C}$) that extends the symmetry to U(3)$_{L}\times$U(3)$_{R}$~\cite{tHooft:1973alw, Gasser:1984gg, Witten:1979vv, Kaiser:2000gs}. Here hadronic decays of $\eta'$ mesons act as a probe for testing the validity and effectiveness of the ChPT extensions and other theoretical models. The $\eta'\to \pi\pi\eta$ decay is particularly suitable for these studies as final-state interactions between $\pi\pi$ and $\pi\eta$ are mainly dominated by scalar contributions, whereas G-parity conservation suppresses vector resonances. Therefore, the properties of the lowest-lying scalar resonances, $f_{0}(500)$ and $a_{0}(980)$, could, in principle, be studied through the $\eta'\to \pi\pi\eta$ decay~\cite{Escribano:2010wt}.

\subsection{Dalitz plot}
The density of the $\eta'\to \pi_1\pi_2\eta$ Dalitz plot, with
$\pi_1\pi_2$ being either the $\pi^+\pi^-$ or $\pi^0\pi^0$ pair,
is completely described by the matrix element of the corresponding
three-body decay amplitude.
The $\eta'\to \pi_1\pi_2\eta$ Dalitz plot is typically expressed in terms of variables $X$ and $Y$, defined as 
\begin{equation}
 X = \frac{\sqrt{3}}{Q}(T_{\pi_1}-T_{\pi_2}),
 ~~Y = \frac{T_{\eta}}{Q}(\frac{m_{\eta}}{m_{\pi}}+2)-1~.
\label{eqn:XY}
\end{equation}
The observables $T_{\pi_1}$, $T_{\pi_2}$, and $T_{\eta}$ denote the kinetic energies of the two final-state pions and $\eta$ in the $\eta'$ rest frame, and $Q=T_{\eta}+T_{\pi_1}+T_{\pi_2}=m_{\eta'}-m_{\eta}-2m_{\pi}$.
 Typically, in recent measurements~\cite{Blik:2009zz, Dorofeev:2006fb, Ablikim:2010kp}, the following
 parametrization
\begin{equation}
 	|M|^2 \sim 1 + aY + bY^2 + cX + dX^2,
\label{eqn:M1sq}
\end{equation}
 (with $a$, $b$, $c$, and $d$ being real-valued parameters) was sufficient to describe the experimental Dalitz plots.
From the theoretical point of view, the $cX$ term should be zero for $\eta'\to \pi^0\pi^0\eta$
 as a consequence of the Bose-Einstein symmetry of the $\pi^0\pi^0$ wave function,
 and for $\eta'\to \pi^+\pi^-\eta$ as a consequence of charge-parity conservation.
For the $\eta'\to \pi^0\pi^0\eta$ Dalitz plot filled only with $|X|$ values
(as in Refs.~\cite{Alde:1986nw,Blik:2009zz}), the $cX$ term makes no sense,
as any linear dependence on $X$ is canceled by adding points with the
plot density
of $|1.-cX|$ and $|1.+cX|$ corresponding to coordinates $-X$ and $X$,
respectively.

 Another parametrization, which historically~\cite{Kalbfleisch:1974ku} was considered in previous
 experiments, is
\begin{equation}
 |M|^2 \sim |1 + \alpha Y|^2 + dX^2,
\label{eqn:M2sq}
\end{equation}
 assuming a linear Y dependence of the $\eta'$ decay amplitude
 with a complex-valued parameter $\alpha$. Because Eq.~(\ref{eqn:M2sq})
 can be transformed into Eq.~(\ref{eqn:M1sq}) via 
 $a = 2\mathrm{Re}(\alpha)$ and $b = \mathrm{Re}^2(\alpha) + \mathrm{Im}^2(\alpha)$,
 the linear parametrization of Y constraints parameter $b$
 from being negative, which is in contradiction
 with all experimental results obtained
 for parameter $b$ so far~\cite{Blik:2009zz, Dorofeev:2006fb, Ablikim:2010kp}.

\subsection{Previous measurements}

\begin{figure*}
\includegraphics[width=1.0\textwidth]{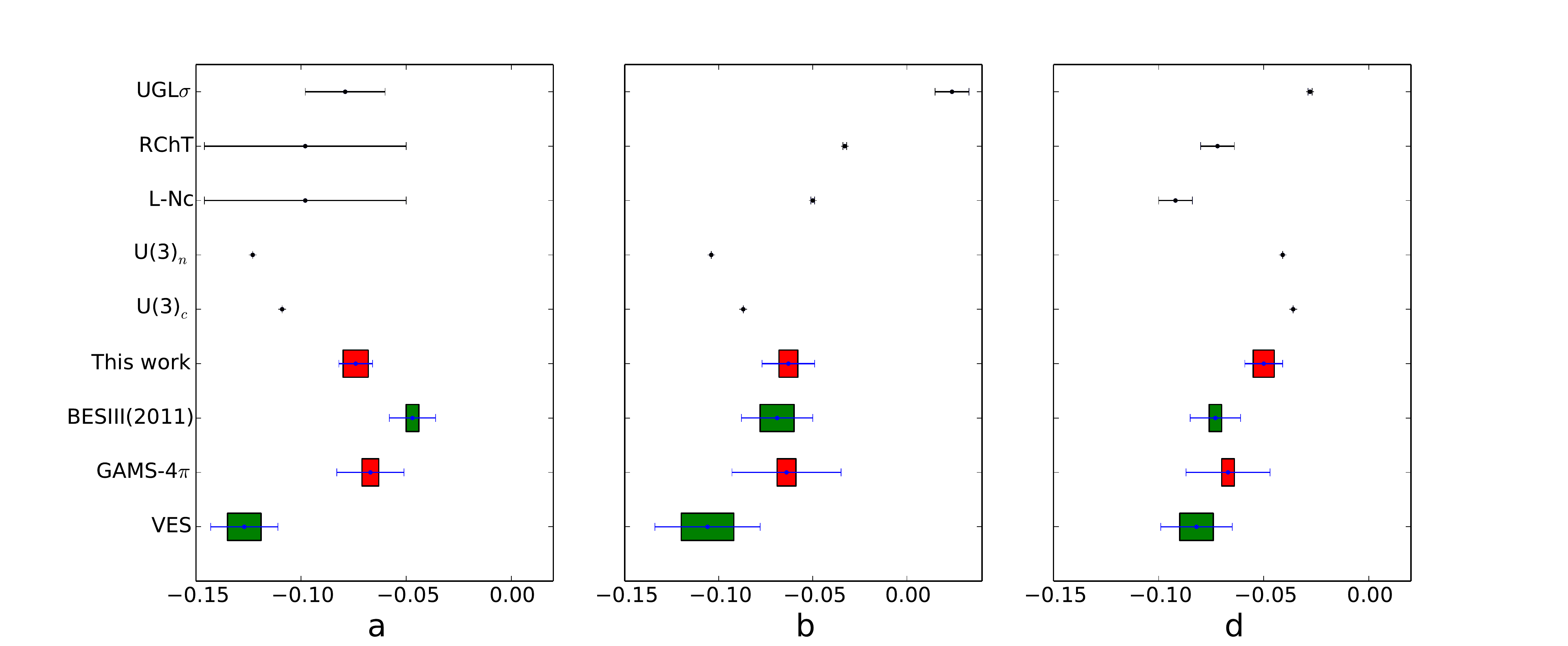}
\caption{ Experimental results and theoretical calculations for the $\eta'\to\pi\pi\eta$ matrix-element parameters,
 with statistical and systematic uncertainties for experimental data points plotted by the blue error bars and boxes, respectively. The $\eta'\to\pi^0\pi^0\eta$ results from this work and by GAMS-4$\pi$~\protect\cite{Blik:2009zz} are shown with red boxes and the $\eta'\to\pi^+\pi^-\eta$ results by VES~\protect\cite{Dorofeev:2006fb} and BESIII~\protect\cite{Ablikim:2010kp} with green.
 The theoretical calculations are shown by black crosses. For Ref.~\protect\cite{Borasoy:2006uv}, they are denoted as
 $U(3)_{n}$ and $U(3)_c$ for the neutral and charged decay mode, respectively. The predictions from Ref.~\protect\cite{Escribano:2010wt} are denoted as $L-N_C$ and RChT for large-$N_C$ ChPT and Resonance Chiral Theory, respectively. The calculation from Ref.~\protect\cite{Fariborz:2014gsa} is denoted as UGL$\sigma$.}
 \label{fig:DPsummary}
\end{figure*}

The results of previous measurements for the standard parametrization of the $\eta'\to\pi\pi\eta$ matrix element with Eq.~(\ref{eqn:M1sq}) are plotted in Fig.~\ref{fig:DPsummary}, along with the most recent theoretical calculations.
For convenience, the results of this work are added in Fig.~\ref{fig:DPsummary} as well.
So far, the neutral decay mode $\eta'\to\pi^0\pi^0\eta$ was only measured with the GAMS spectrometers~\cite{Alde:1986nw,Blik:2009zz}. The first experimental data were taken with the GAMS-2000
spectrometer at the IHEP accelerator U-70~\cite{Alde:1986nw}, where
$\eta'$ mesons were produced in the charge-exchange reaction
$\pi^{-}p\to\eta' n$, with $5.4\cdot10^3$
$\eta'\to\pi^0\pi^0\eta\to6\gamma$ decays accumulated. A more comprehensive study of this decay was later made with the upgraded GAMS-2000 spectrometer, GAMS-4$\pi$, which accumulated $1.5\cdot10^4$ decays~\cite{Blik:2009zz}.
The decay with charged pions, $\eta'\to \pi^+\pi^-\eta$, was measured by the CLEO~\cite{Briere:1999bp}, VES~\cite{Dorofeev:2006fb} and BESIII~\cite{Ablikim:2010kp} collaborations. For their analyses, CLEO, VES, and  BESIII accumulated $6.7\cdot10^3$, $20.1\cdot10^3$, and $43.8\cdot10^3$ $\eta'\to \pi^+\pi^-\eta$ decays, respectively.
Compared to the other experiments, the measurement by CLEO~\cite{Briere:1999bp} only determined the linear parametrization in Eq.~(\ref{eqn:M2sq}). Comparison of the recent results from BESIII~\cite{Ablikim:2010kp} and GAMS-4$\pi$~\cite{Blik:2009zz} indicates that the isospin limit is a good approximation. 
The latest, high-statistics results from BESIII~\cite{Ablikim:2017irx} for both the decay modes, which are based on the same data sample that was used to analyze the $\eta'\to\pi\pi\pi$ decay~\cite{Ablikim:2016frj}, appeared after the submission of this work and will, therefore, not be considered in the present paper.

\subsection{Theoretical calculations}
There are a few recent theoretical studies of $\eta'\to\pi\pi\eta$ decays. One of the studies was made within the framework of U(3) chiral effective-field theory in combination with a relativistic coupled-channels approach~\cite{Beisert:2003zs, Borasoy:2005du, Borasoy:2006uv}, using the VES results and parameter $a$ from GAMS-4$\pi$ in their fitting procedure. The results of Ref.~\cite{Borasoy:2006uv} are shown in Fig.~\ref{fig:DPsummary} as $U(3)_{n}$ and $U(3)_{c}$ for the $\eta'\to\pi^0\pi^0\eta$ and $\eta'\to\pi^+\pi^-\eta$ decays, respectively. 

Another study was conducted within the large-$N_C$ ChPT, at lowest and next-to-leading orders, and Resonance Chiral Theory (RChT) in the leading $1/N_C$ approximation, with higher-order effects, such as $\pi\pi$ final-state interactions, being taken into account through a detailed unitarization procedure~\cite{Escribano:2010wt}. In the isospin limit, the intrinsic calculations for the charged and neutral decay amplitudes are the same. For the unitarization procedure that works within large-$N_C$ ChPT and RChT, experimental results are needed to fix particular theoretical constants. If one takes parameter $a$ as its average from Refs.~\cite{Blik:2009zz, Dorofeev:2006fb} and the uncertainty in $a$ as the
absolute difference in the two results, with their own uncertainties added, then the large-$N_C$ ChPT predicts $b=-0.050(1)$ and $d=-0.092(8)$, and RChT gives $b=-0.033(1)$ and $d=-0.072(8)$. The latter predictions are in slightly better agreement with the BESIII and GAMS-4$\pi$ results, compared to the calculations with the U(3) chiral effective-field theory~\cite{Borasoy:2006uv}. In addition to the standard parametrization of $\eta'\to\pi\pi\eta$, an expanded parametrization that includes two additional higher-order terms was also checked in Ref.~\cite{Escribano:2010wt}:
\begin{equation}
 |M|^2 \sim 1 + aY + bY^2 + dX^2 + \kappa_{21}YX^2 + \kappa_{40}X^4,
\label{eqn:M3sq}
\end{equation}
where the notation of parameters $\kappa_{21}$ and $\kappa_{40}$ is left as in Ref.~\cite{Escribano:2010wt}. Such a parametrization was motivated by the fact that, in the chiral limit, expansions of $Y$ should be of similar magnitude to those of order $X^2$, with the expected hierarchy of the parameters to be $|a| \sim |d| \gg |b| \sim |\kappa_{21}| \sim |\kappa_{40}|$, which was not confirmed by existing data, giving $|a| \sim |b| \sim |d|$. For these additional terms, large-$N_C$ ChPT predicts $\kappa_{21}=0.003(2)$ and $\kappa_{40}=0.002(1)$, and RChT gives $\kappa_{21}=-0.009(2)$ and $\kappa_{40}=0.001(1)$.

One more calculation of the $\eta'\to\pi\pi\eta$ decay, which includes mixing between scalar mesons, was studied within a generalized linear sigma model containing two nonets of scalar mesons and two nonets of pseudoscalar mesons~\cite{Fariborz:2014gsa}. The four nonets are two quark-antiquark and two four-quark states, respectively. For this calculation, denoted as UGL$\sigma$ in Fig.~\ref{fig:DPsummary}, agreement with the experimental data was obtained only for parameter $a$, but not for $b$ or $d$.

 The main difference between the $\eta'\to \pi^0\pi^0\eta$ and $\eta'\to \pi^+\pi^-\eta$ decays, which is caused
 by strong final-state interactions, could especially be visible in the $\eta'\to \pi^0\pi^0\eta$ spectrum
 below the $\pi^+\pi^-$ threshold, where a pronounced cusp is expected.
 The contributions from the final-state interactions for the $\eta'\to\pi\pi\eta$ decays were calculated within the framework of non-relativistic effective-field theory (NREFT)~\cite{Kubis:2009sb}, which was developed and successfully used for extracting $\pi\pi$ scattering lengths from $K \to 3\pi$ decays~\cite{Colangelo:2006va, Bissegger:2007yq, Bissegger:2008ff}. Based on the previously determined scattering lengths, the expected magnitude of the cusp is 6\%~\cite{Kubis_privat} (was 8\% in the original work~\cite{Kubis:2009sb}). To check the reliability of such a prediction, a new measurement of $\eta'\to\pi^0\pi^0\eta$ with good statistical
accuracy and experimental resolution is needed.

 The most recent dispersive analysis of the $\eta'\to\pi\pi\eta$ decay amplitude, which is based on the fundamental principles of analyticity and unitarity, is presented in Ref.~\cite{Isken:2017dkw}. 
In the analysis framework, final-state interactions are fully taken into account, and the dispersive representation
 relies only on input for the $\pi\pi$ and $\pi\eta$ scattering phase shifts.
 Because the dispersion relation contains subtraction constants that cannot be fixed by unitarity, these parameters
 were determined by fitting existing Dalitz-plot data. The prediction of a low-energy theorem was studied and
 the dispersive fit was compared to variants of ChPT.
  
This work presents an experimental study of the $\eta'\to\pi^{0}\pi^{0}\eta$ amplitude based on an analysis of $\sim 1.2\cdot 10^5$ $\eta'\to\pi^0\pi^0\eta$ decays. In addition to the determination of the standard matrix-element parameters $a$, $b$, and $d$, the sensitivity to higher-order terms $\kappa_{21}YX^2$ and $\kappa_{40}X^4$, suggested in Ref.~\cite{Escribano:2010wt}, is tested. The magnitude of the cusp effect at the $\pi^+\pi^-$ threshold is checked within the NREFT framework~\cite{Kubis:2009sb} by using $\pi\pi$ scattering lengths extracted from $K\to3\pi$ decays. Acceptance-corrected Dalitz plots and ratios of the $X$, $Y$, $m(\pi^0\pi^0)$, and $m(\pi^0\eta)$ distributions to phase space are also provided, which could be used in theoretical approaches that determine some of their parameters from fitting to experimental data.

\section{Experimental setup}
\label{sec:Setup}
\begin{figure}
\includegraphics[width=0.8\linewidth]{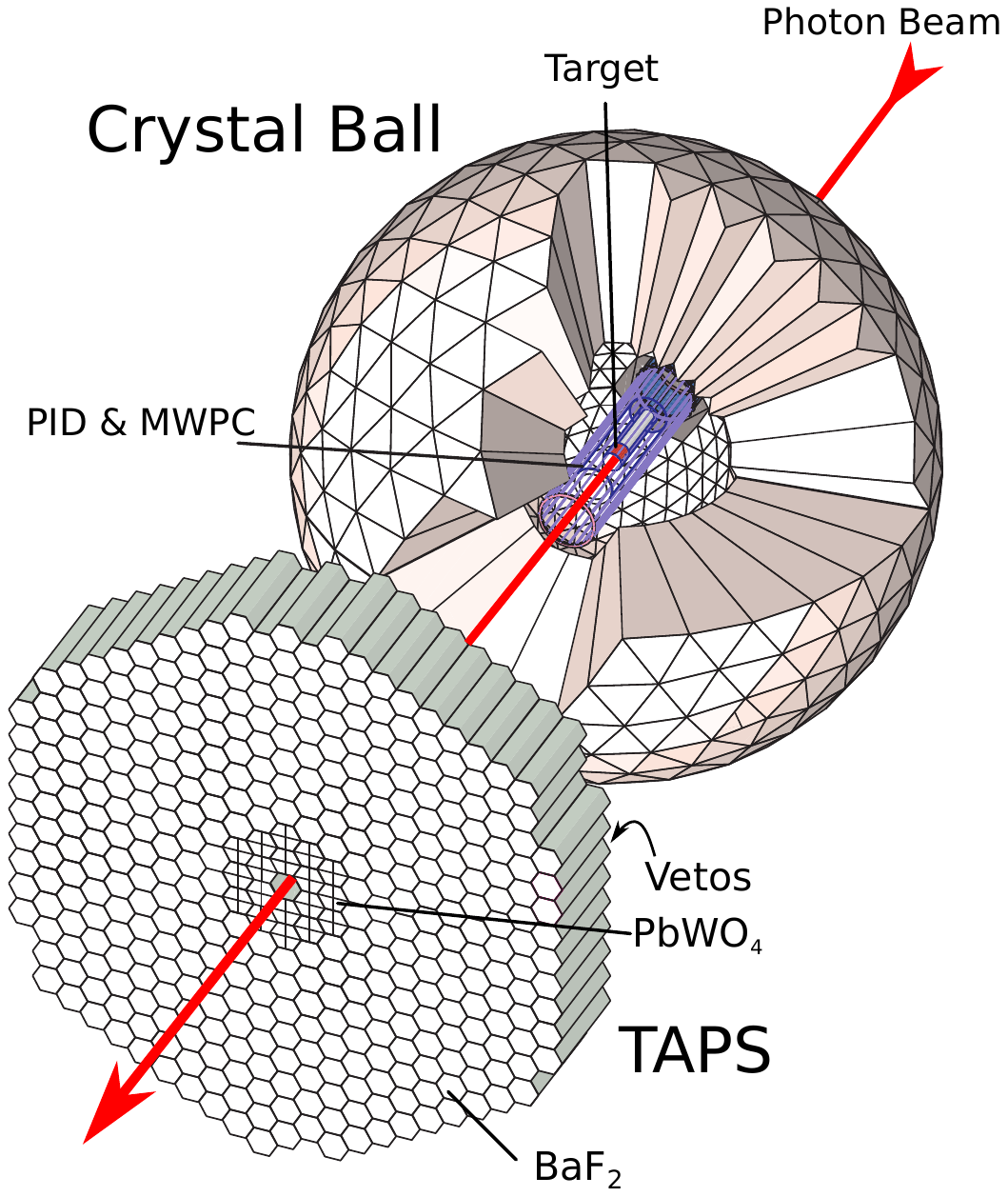}
\caption{A general sketch of the detectors used in the A2 experimental setup.}
 \label{fig:CBTAPS}
\end{figure}
The process $\gamma p\to\eta'p\to\pi^0\pi^0\eta p\to6\gamma p$
was measured with the A2 experimental setup using the Crystal Ball (CB)~\cite{Starostin:2001zz} as a central
calorimeter and TAPS~\cite{Novotny:1991ht, Gabler:1994ay} as a forward calorimeter.
These detectors were installed in the energy-tagged bremsstrahlung photon beam of
the Mainz Microtron (MAMI)~\cite{Herminghaus:1983nv, Kaiser:2008zza}.
The photon energies are determined by the Mainz end-point tagger (EPT)~\cite{Adlarson:2015byy}.

The CB detector is a sphere consisting of 672 optically isolated NaI(Tl) crystals, shaped as truncated triangular pyramids, pointing toward the center of the sphere. The crystals are arranged in two hemispheres that cover 93\% of $4\pi$ sr, sitting outside a central spherical cavity with a radius of 25~cm, which is designed to hold the target and inner detectors. In the A2 experiments from 2007, TAPS was initially arranged in a plane consisting of 384 BaF$_2$ counters of hexagonal cross section. It was installed 1.45~m downstream from the CB center covering the full azimuthal range for polar angles from $1^\circ$ to $20^\circ$. Later on, 18 BaF$_2$ crystals, covering polar angles from $1^\circ$ to $5^\circ$, were replaced with 72 PbWO$_4$ crystals. This allowed running with a high MAMI electron current without decreasing the TAPS efficiency due to the high count rate in the crystals close to the photon-beam line. More details on the calorimeters and their resolutions are given in Ref.~\cite{Prakhov:2008ff} and references therein.

 The target is surrounded by a Particle IDentification
 (PID) detector~\footnote{D.~Watts, {\it Proceedings of the 11th International Conference on Calorimetry in Particle
 Physics}, Perugia, Italy, 2004 (World Scientific, Singapore, 2005), p. 560.} used to distinguish between charged and
 neutral particles. It is made of 24 scintillator bars
 (50 cm long, 4 mm thick) arranged as a cylinder with a radius of 12 cm.
 A general sketch of the A2 setup is shown
 in Fig.~\ref{fig:CBTAPS}.
 A charged particle tracker, MWPC, also shown in this figure
 (consisting of two cylindrical multi-wire proportional chambers inside each other),
 was not used in the present measurement.

The present measurement was conducted in 2014 with a 1604-MeV unpolarized electron beam from the Mainz Microtron, MAMI C~\cite{Herminghaus:1983nv, Kaiser:2008zza}.  Bremsstrahlung photons, produced by the beam electrons
 in a 10-$\mu$m Cu radiator and collimated by a 4-mm-diameter Pb collimator, impinged on a 10-cm-long liquid hydrogen (LH$_2$) target located in the center of the CB.
 The energies of the incident photons were analyzed from 1426~MeV up to 1577~MeV by detecting the postbremsstrahlung electrons in 47 focal-plane detectors of the EPT magnetic spectrometer.
 It was especially built to conduct $\eta'$ measurements by covering the low-energy range of postbremsstrahlung electrons. The uncertainty of the EPT in $E_\gamma$ due to the width of its focal-plane detectors was about $\pm 1.6$~MeV, with a similar value (i.e., $\sim 1.6$~MeV) in the systematic uncertainty in $E_\gamma$ due to the EPT energy calibration. The energy calibration of the EPT is based only on the simulation of electron tracing, using measured magnetic-field maps. The correctness of this calibration, as well as its uncertainty, was checked by measuring the position of the $\eta'$ threshold, $E_\gamma \approx 1447$~MeV.

Because the EPT experiments were mainly dedicated to studying $\eta'$ physics, the experimental trigger required the total energy deposit
in the CB to be greater than $\sim$540~MeV to suppress triggering on reactions below the $\eta'$ mass threshold. \\

\section{Event Selection}
\label{sec:EventSelection}
The process $\gamma p\to\eta'p\to\pi^0\pi^0\eta p\to 6\gamma p$ was searched for in events with seven energy-deposit clusters detected in both the CB and TAPS within the trigger time window, assuming that one of the seven clusters was from the recoil proton. The offline cluster algorithm is optimized for finding a group of adjacent crystals in which the energy is deposited by a single-photon electromagnetic (e/m) shower, but it also works well for recoil protons.
 In the present beam-energy range, the recoil protons from the reaction $\gamma p\to\eta' p$ were produced within the polar-angle range covered solely by TAPS, which required only events with at least one cluster detected in TAPS to be considered. In case of more than one cluster in TAPS, the recoil proton could be identified by a Time-of-Flight (TOF) method, based on the information from TAPS TDCs.
In general, the selection of event candidates and the reconstruction of the reaction kinematics were based on the kinematic-fit technique, which uses a constrained least-squares fit with constraints based on energy and momentum conservation~\cite{blolo}. Typically, such a technique is sufficient to identify which cluster is from the recoil proton.
 Because the LH$_2$ target was 10-cm long, the vertex coordinate $z$ (along the beam line) was used as another
 kinematic-fit variable, which can be used as either its measured or free parameter.
 This improves angular resolution as the kinematic-fit then determines the recoil-particle angles by varying $z$ instead
 of using the cluster angles, which are calculated by assuming $z=0.$
In addition to seven-cluster events, the cases with eight clusters were also checked when the least energetic
 cluster in TAPS had energy below 40~MeV. For process under the study, there are e/m showers leaking from the CB
 through its downstream tunnel to TAPS, and the events with small leakage still can be reconstructed by the kinematic fit, though with poorer resolution. The recoil proton cannot be eliminated with such an approach as their
 kinetic energy is much larger than 40~MeV. In the end, including eight-cluster events in the analysis gained the
 experimental statistic just by 3\%.

 To search for candidates to $\eta'\to\pi^0\pi^0\eta \to 6\gamma$ decays,
 the $\gamma p\to \pi^0\pi^0\eta p\to 6\gamma p$ kinematic-fit hypothesis was checked by using
 the four main constraints (conservation of energy and three momentum projections) and three additional
 constraints on the invariant masses of two neutral pions and $\eta$ decaying into two photons.
 The number of 45 possible combinations to pair six photons to form the three final-state mesons could be decreased
 by checking invariant masses of cluster pairs before testing the corresponding hypothesis. The number of 7 possible
 combinations to pick the proton cluster can be decreased to the number of the clusters detected in TAPS.
 The events for which at least one pairing combination satisfied
 the tested hypothesis at the 1\% confidence level, $\CL$, (i.e., with a probability greater
 than 1\%) were selected for further analysis. The pairing combination with the largest $\CL$ was used
 to reconstruct the reaction kinematics. The combinatorial background from mispairing six photons
 to three mesons was found to be quite small, with further decreasing if tightening a selection criterion on
 the kinematic-fit $\CL$ or adding a constraint on the $\eta'$ mass.
  
The determination of the experimental acceptance for the process $\gamma p\to\eta'p\to\pi^0\pi^0\eta p\to6\gamma p$ and the study of its possible background reactions were based on their Monte Carlo (MC) simulation. All MC events were propagated through a GEANT simulation (version 4.9.6) of the experimental setup. To reproduce the experimental resolutions, the GEANT output was additionally smeared, allowing the simulated and experimental data to be analyzed in the same way. The MC smearing was adjusted to match the experimental invariant-mass resolutions and kinematic-fit confidence level ($\CL$) distributions. The simulated events were also checked whether they passed the trigger requirements.

 The reaction $\gamma p\to\pi^0\pi^0\pi^0p\to6\gamma p$, in which $\pi^0$ mesons are produced via intermediate baryon
 states, was used for the energy calibration of the calorimeters and the determination of their energy resolution.
 This reaction was also used to determine the additional smearing that was needed to reproduce the experimental resolution in the analysis of the MC simulations. The final adjustment for the energy calibration and the experimental
resolution was made by comparing the $\eta'\to\pi^0\pi^0\eta\to6\gamma$ peak in the analysis of the
experimental data and the corresponding MC simulation.
 
Because the experimental acceptance and the invariant-mass resolution depend on the $\eta'$ production angle, systematic differences can appear if the simulated data do not reflect the actual production distributions. To diminish such systematic effects, the reaction $\gamma p\to\eta' p$ was generated according to its actual yield and angular spectra measured as a function of energy in the same experiment~\cite{Kashevarov:2017kqb}. In that measurement, the $\gamma p\to\eta' p$ differential cross sections obtained
from the analysis of $\eta'\to \pi^0\pi^0\eta\to 6\gamma$ and $\eta'\to \gamma\gamma$ decays were found to be in good agreement, which confirmed the data quality and the analysis reliability. To diminish the impact of the experimental resolution on the shape of the
 acceptance-corrected Dalitz plot, the detection efficiency for it was determined from the MC simulation in which the $\eta'\to\pi^0\pi^0\eta$ decay was generated close to the actual density distribution of the experimental plot,
initially obtained by using the MC simulation with the $\eta'\to \pi^0\pi^0\eta$ decay generated as phase space.

The study of possible background reactions via their MC simulation showed that the process $\gamma p\to3\pi^0p\to6\gamma p$ can mimic $\gamma p\to\pi^0\pi^0\eta p\to6\gamma p$ via the combinatorial mismatching of six photons to three $\pi^0$,
which can occur when the invariant mass $m(6\gamma)>820$~MeV. Similarly, the process $\gamma p\to\pi^0\eta p\to4\pi^0p\to8\gamma p$ can mimic the signal channel when two of the eight final-state photons are not detected. The suppression of the $\gamma p\to3\pi^0 p\to6\gamma p$ background was done by testing the corresponding kinematic-fit hypothesis and rejecting events
based on its $\CL$, which also resulted in some losses of the actual $\gamma p\to \pi^0\pi^0\eta p\to6\gamma p$ events. The suppression of the $\gamma p\to\pi^0\eta p\to4\pi^0 p\to 8\gamma p$ background can be achieved solely by tightening the cut on the $\CL$ of the $\gamma p\to\pi^0\pi^0\eta p\to6\gamma p$ hypothesis itself.

In addition to background from other reactions, there are two more background sources. The first comes from interactions of the beam photons with the target cell windows. This background can be studied by analyzing data samples with an empty target cell,
i.e., containing no liquid hydrogen. The contribution from such events was found to be very small and was neglected in the further analysis. A second background is caused by random coincidences of the tagger counts with the experimental trigger. This background was subtracted directly from the experimental spectra by using event samples with only random tagger coincidences, which is a standard technique for experiments using a beam of tagged bremsstrahlung photons~\cite{Owens:1990du, Prakhov:2008ff, McNicoll:2010qk}.
\begin{figure*}
\includegraphics[width=1.0\textwidth]{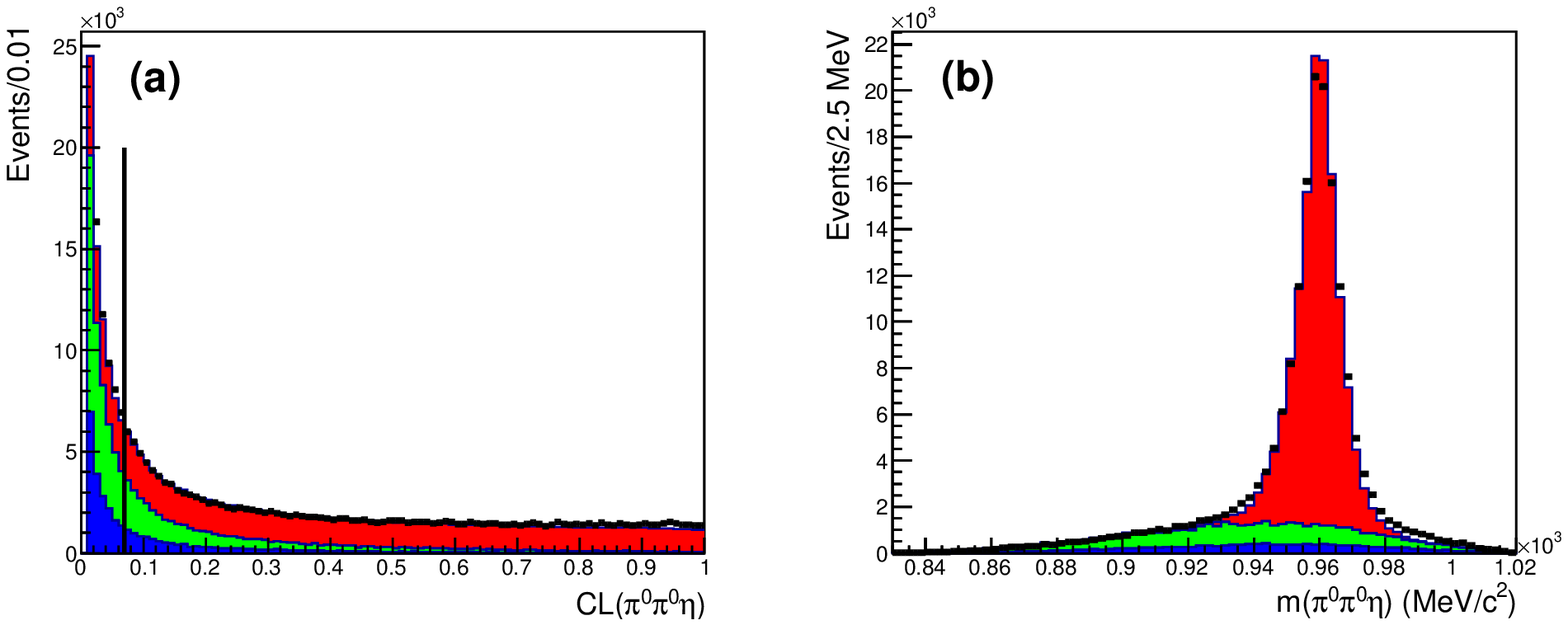}
\caption{
 (a) Probability (or $\CL$) distributions from testing the $\gamma p\to\pi^0\pi^0\eta p\to6\gamma p$ hypothesis
 for the experimental data (black points) and the sum of MC simulations for $\gamma p\to\eta'p\to\pi^0\pi^0\eta p\to6\gamma p$, $\gamma p\to\pi^0\pi^0\pi^0 p\to6\gamma p$ , and $\gamma p\to\pi^0\eta p\to4\pi^0 p\to 8\gamma p$, shown
 by the red, blue, and green areas, respectively. All events also passed the requirement 
 $\CL(\gamma p \to\pi^0\pi^0\pi^0p\to 6\gamma p)<0.08$~.
 (b) Invariant-mass $m(\pi^0\pi^0\eta)$ distributions for the same samples after passing an additional
  requirement $\CL(\gamma p\to\pi^0\pi^0\eta p\to6\gamma p)>0.07$, shown by the vertical line in (a).
}
 \label{fig:Fig_PDF_Meta2pi0}
\end{figure*}

The event-selection procedure included tests of several kinematic-fit hypotheses.
First, all events had to pass the $\gamma p \to 6\gamma p$ hypothesis with its $\CL>0.01$.
As mentioned above, to test the $\gamma p\to\pi^0\pi^0\eta p\to6\gamma p$ hypothesis, 45 combinations of pairing
 the six photons to two $\pi^0$ and one $\eta$ meson were checked, and the combination with the best $\CL$ was
 selected for further analysis. Similarly, for the $\gamma p\to\pi^0\pi^0\pi^0 p\to6\gamma p$ hypothesis,
 the combination with the best $\CL$, from 15 possible pairings of the six photons to three $\pi^0$ mesons, was selected.
 The probability (or $\CL$) distribution for experimental events selected as $\gamma p\to\pi^0\pi^0\eta p\to6\gamma p$
 with $\CL>0.01$ is shown in Fig.~\ref{fig:Fig_PDF_Meta2pi0}(a), after partially suppressing the $\gamma p\to\pi^0\pi^0\pi^0 p\to6\gamma p$ background by requiring $\CL(\pi^{0}\pi^{0}\pi^{0})<0.08$~.
This experimental distribution is described well by the sum of similar probability distributions for the MC simulations
of $\gamma p\to\eta'p\to\pi^0\pi^0\eta p\to6\gamma p$, $\gamma p\to\pi^0\pi^0\pi^0 p\to6\gamma p$, and 
$\gamma p\to\pi^0\eta p\to4\pi^0 p\to 8\gamma p$, the events of which were selected with the same criteria.
 For further suppression of the background reactions, only events with $\CL(\pi^{0}\pi^{0}\eta)>0.07$, shown by
 the vertical line in Fig.~\ref{fig:Fig_PDF_Meta2pi0}(a), were selected for the final analysis.
  The invariant-mass $m(\pi^0\pi^0\eta)$ distributions for the selected experimental and MC events are depicted
 in Fig.~\ref{fig:Fig_PDF_Meta2pi0}(b), which shows the level of the background contributions remaining
 under the $\eta'$ peak. Because fitting the $\gamma p\to\eta'p\to\pi^0\pi^0\eta p\to6\gamma p$ hypothesis with
 an additional constraint on the $\eta'$ mass does not eliminate the background under the $\eta'$ peak,
 the experimental number of $\eta'$ mesons in every individual bin of various distributions was then determined by
 fitting the $m(\pi^0\pi^0\eta)$ spectra corresponding to those bins.
\begin{figure*}
\includegraphics[width=0.48\textwidth]{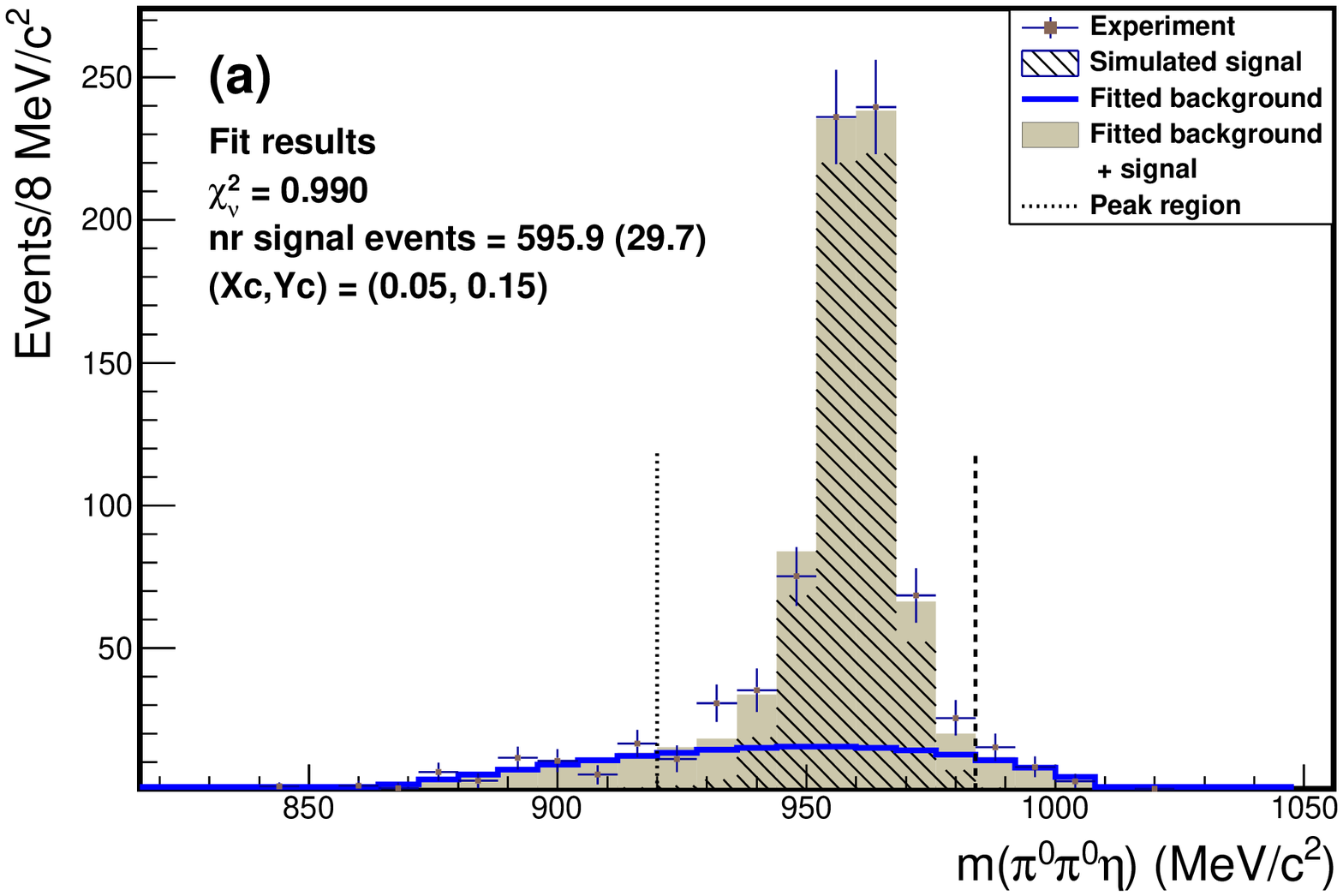}
\includegraphics[width=0.48\textwidth]{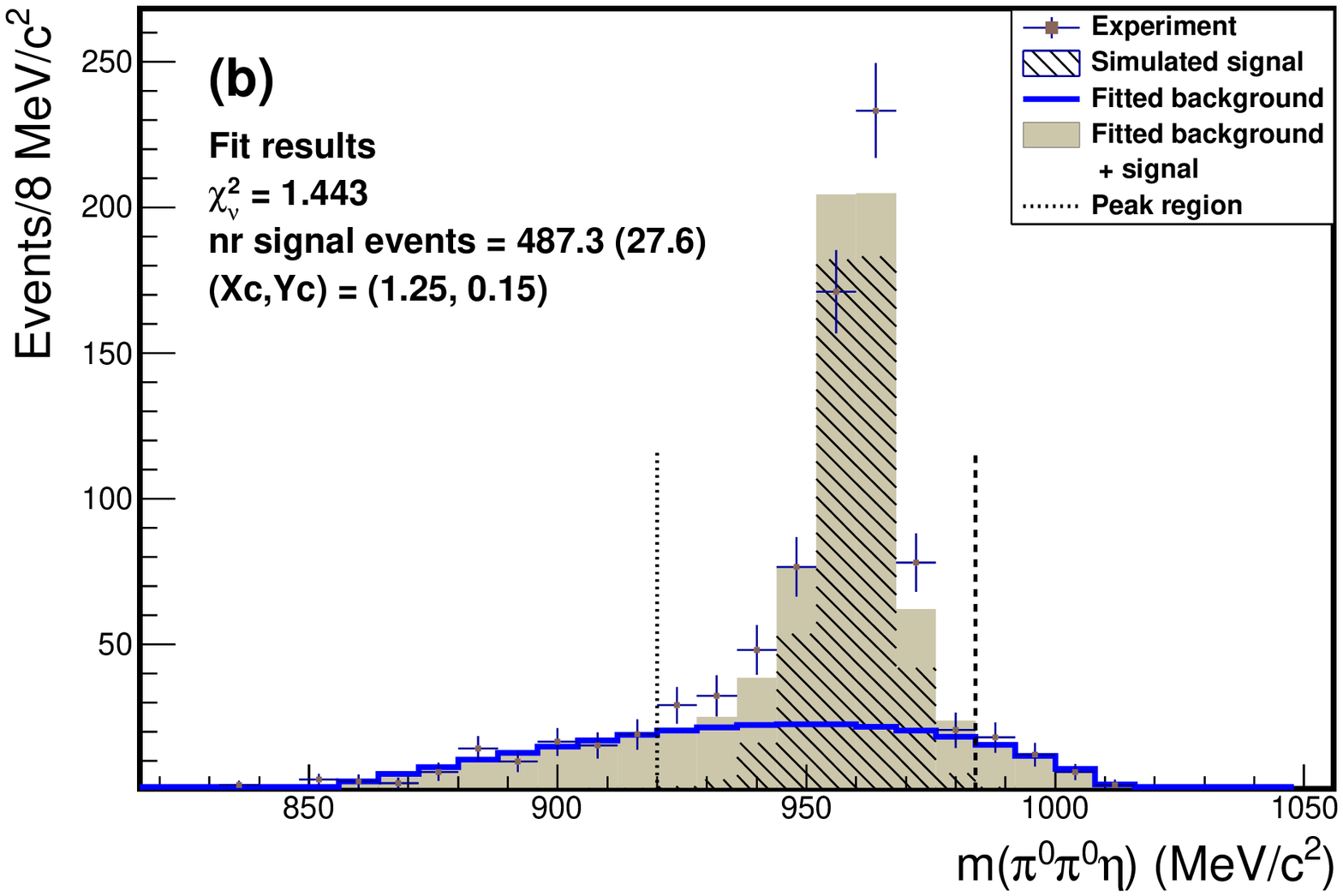}
\caption{Fitting procedure to measure the number of $\eta'\to\pi^0\pi^0\eta$ decays in the $m(\pi^0\pi^0\eta)$ spectra
 corresponding to two different bins of the Dalitz plot: (a) with bin centers at ($Xc$, $Yc$) = (0.05, 0.15) and 
 (b) ($Xc$, $Yc$) = (1.25, 0.15). The bin width is 0.1 for both $X$ and $Y$.
 The experimental spectra (crosses) are fitted with the sum (shaded in gray) of the signal line shape, fixed from
 its MC simulation (hatched in black), and a polynomial of order 4 for the background (blue line), binned as 
 a histogram of the same bin size.
 The experimental number of $\eta'$ decays is determined by integrating the signal line shape from the fit
 within $920<m(\pi^0\pi^0\eta)<980$~MeV (vertical dashed lines).}
 \label{fig:fitexamples}
\end{figure*}

 This fitting procedure is illustrated in Fig.~\ref{fig:fitexamples} for two different bins of the $\eta'\to \pi^0\pi^0\eta$ Dalitz plot. To determine the number of measured $\eta'$ mesons, the experimental $m(\pi^0\pi^0\eta)$ distribution
 was fitted with the sum of the signal line shape, fixed from a high-statistics $\gamma p\to\eta'p\to\pi^0\pi^0\eta p\to6\gamma p$ MC sample, and a polynomial of order 4 for describing the background. The measured number of $\eta'$ mesons in a given bin
was determined from the number of events in the corresponding MC distribution for $\eta'\to \pi^0\pi^0\eta$ decays
times the weight factor obtained from the fit. The uncertainty in the number of measured $\eta'$ in a given bin was based on the uncertainties in the signal MC distribution,
the uncertainty in its weight factor from the fit, and the polynomial-fit uncertainty. The total number of measured $\eta'$ decays integrated over the entire Dalitz plot
 was obtained as $\sim1.241(4)(8)\cdot10^{5}$, with the first uncertainty being determined by the experimental
 statistics and the second being the systematic uncertainty due to the fitting method.
 The latter uncertainty was estimated in two ways. First, the same fitting procedure was apllied to the sum of MC simulations for the signal and background contributions, with the number of signal events taken as experimentally observed.
Second, the number of $\eta'$ decays was obtained by subtracting the background from the experimental spectra,
 based on the fit results (as shown in Fig.~\ref{fig:fitexamples}) for polynomial describing the background distribution. 

 The acceptance-corrected Dalitz plot can be obtained by correcting the measured number of $\eta'$ decays in each bin
 by the ratio of the reconstructed and generated events from the $\gamma p\to\eta'p\to\pi^0\pi^0\eta p\to6\gamma p$ MC simulation in the given bin. Figure~\ref{fig:PatrikDP}(a) shows such a plot for the bin width 0.1 in both $X$ and $Y$,
 which allows sufficient statistics for a reliable determination of the number of $\eta'$ decays in each bin.
 Based on the analysis of the MC simulation for $\eta'\to \pi^0\pi^0\eta$ decays, the Dalitz-plot acceptance was found to be
 almost uniform (with the averaged efficiency of 24.5\%), but with its boundaries smeared by the experimental resolution,
 the average of which for $X$ and $Y$ was determined as $\delta X=0.09$ and $\delta Y=0.076$~.
 Because the $\eta'$ mass constraint was not used to obtain the background-free Dalitz plot, the reconstructed events can migrate outside the physical boundaries of the plot determined by the $\eta'$ mass.
 To determine the $\eta'\to \pi^0\pi^0\eta$ matrix-element parameters, only bins that are fully inside the Dalitz plot's physical boundaries were considered. These bins contained, for the given bin size, 1.13$\cdot10^5$ measured $\eta'$ decays.

 As a cross check of the analysis discussed above (Analysis I), an independent analysis (Analysis II) of the same data
 was performed to measure the $\eta'\to \pi^0\pi^0\eta$ Dalitz plot by using the $\eta'$ mass as a constraint
 in the kinematic-fit. The main advantage of using the $\eta'$ mass constraint is in the
possibility of creating a sample of $\eta'\to \pi^0\pi^0\eta$ decays, which allows fitting on
an event-by-event basis, and in improving resolution in measured observables.
 Also, all events are reconstructed within the $\eta'\to \pi^0\pi^0\eta$ physical boundaries.
 The main disadvantage of this approach is the background contributions from $\gamma p\to\pi^0\pi^0\pi^0 p\to6\gamma p$
 and $\gamma p\to\pi^0\eta p\to4\pi^0 p\to 8\gamma p$ remaining in the experimental data sample.
 Although such background could be suppressed by tightening selection criteria, this would also decrease the number of $\eta'\to \pi^0\pi^0\eta$ decays available for the analysis. 
  
 The main experimental details of Analysis II are given in Ref.~\cite{Kashevarov:2017kqb}.
 That work included the measurement of the $\gamma p \to \eta' p$ differential cross sections by using
 $\eta'\to \pi^0\pi^0\eta\to 6\gamma$ and $\eta'\to \gamma\gamma$ decays with the method similar to Analysis I,
 i.e, by fitting the $\eta'$ signal above background in bins of differential cross sections.
  Such an analysis tests the $\gamma p\to \pi^0\pi^0\eta p\to 6\gamma p$ kinematic-fit hypothesis,
  involving three additional constraints on the invariant masses of two neutral pions and $\eta$ decaying
  into two photons.
 To measure the $\eta'\to \pi^0\pi^0\eta$ Dalitz plot, the $\gamma p\to\eta'p\to\pi^0\pi^0\eta p\to6\gamma p$
 kinematic-fit hypothesis (with the fourth additional constraint on the invariant mass of
 the two neutral pions and $\eta$ to equal the $\eta'$ mass) was also tested.
 The experimental sample of $\sim1.23\cdot10^5$ $\eta'\to \pi^0\pi^0\eta\to 6\gamma$ decays, reconstructed by
 the fit with the four additional constraints, was then selected by requiring
 $\CL(\gamma p\to \eta' p\to \pi^0\pi^0\eta p\to 6\gamma p)>0.04$ for the main hypothesis
 and $\CL(\gamma p\to\pi^{0}\pi^{0}\pi^{0}p)<0.0075$ for the background hypothesis (involving
 three additional constraints on the invariant masses of the three neutral pions).
 The resolution in $X$ and $Y$ for such selection criteria was determines as 0.07 and 0.06, respectively.
\begin{figure*}
\includegraphics[width=1.0\textwidth]{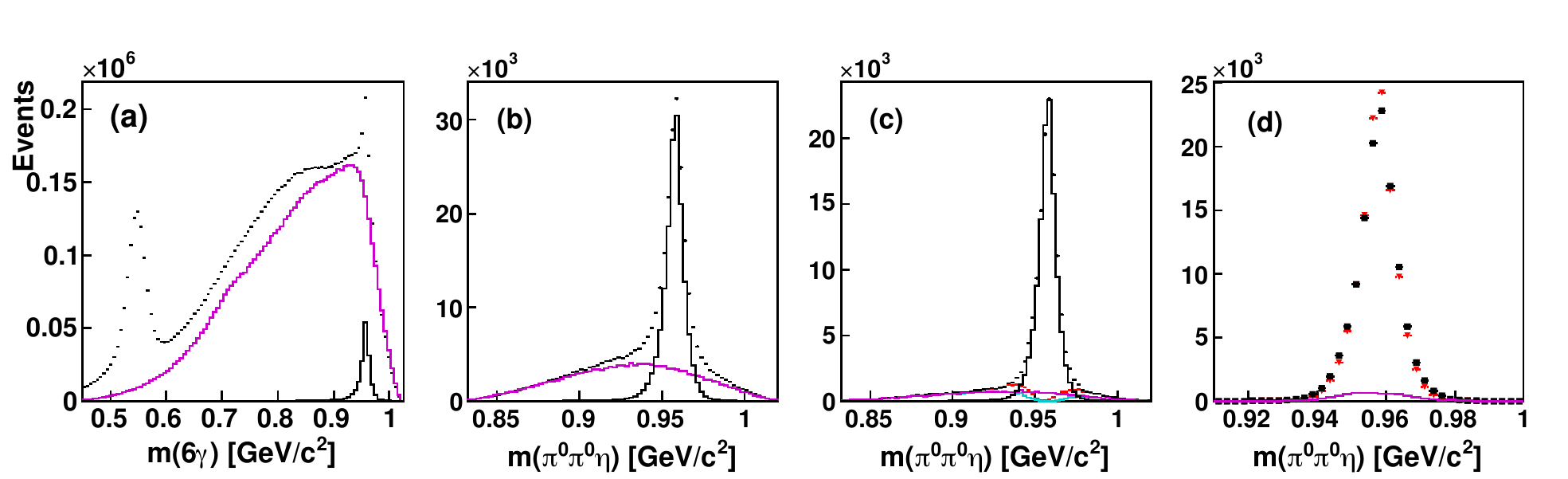}
\caption{
 Experimental $m(6\gamma)$ and $m(\pi^0\pi^0\eta\to 6\gamma)$ invariant-mass distributions (black points)
 from Analysis II compared to various MC simulations, with the combined background from
 $\gamma p\to\pi^0\pi^0\pi^0 p\to6\gamma p$ and $\gamma p\to\pi^0\eta p\to4\pi^0 p\to 8\gamma p$ shown
 by the solid magenta line and the $\gamma p\to \eta' p\to \pi^0\pi^0\eta p\to 6\gamma p$
 MC simulation shown in (a), (b), and (c) by the solid black line:
 (a)~events reconstructed by testing the $\gamma p\to 6\gamma p$ hypothesis (no additional constraints on
 invariant masses of the final-state photons) and selected with the corresponding CL$>$1\%;
 (b)~events reconstructed by testing the $\gamma p\to \pi^0\pi^0\eta p\to 6\gamma p$ hypothesis
  (with the three additional constraints on the photons' invariant masses) and selected with the corresponding CL$>$1\%;
 (c)~events reconstructed as in (b), but selected by requiring
    CL$(\gamma p\to \pi^0\pi^0\eta p\to 6\gamma p)>4\%$ and CL$(\gamma p\to 3\pi^0 p\to 6\gamma p)<0.75\%$;
    events for which CL$(\gamma p\to \eta' p\to \pi^0\pi^0\eta p\to 6\gamma p)<4\%$ from the fit
    with the fourth additional constraint on the $\eta'$ mass are shown by red points for experimental data
    and by the cyan line for the combined background;
 (d)~events reconstructed and selected as in (c), but for which
     CL$(\gamma p\to \eta' p\to \pi^0\pi^0\eta p\to 6\gamma p)>4\%$,
     with the $\gamma p\to \eta' p\to \pi^0\pi^0\eta p\to 6\gamma p$
     MC simulation shown by red points.
}
 \label{fig:etapr_eta2pi0_ept2014_4x1}
\end{figure*}
 The background-estimation procedure of Analysis II is illustrated in
 Fig.~\ref{fig:etapr_eta2pi0_ept2014_4x1}.
 Invariant-mass distributions, $m(6\gamma)$,
 for events reconstructed by testing $\gamma p\to 6\gamma p$ hypothesis, without any additional constraints
 on invariant masses, and selected by requiring the corresponding CL$>$1\%
 are shown in Fig.~\ref{fig:etapr_eta2pi0_ept2014_4x1}(a)
 for the experimental data, the MC simulations of the signal channel
 $\gamma p\to \eta' p\to \pi^0\pi^0\eta p\to 6\gamma p$
 and the background channel $\gamma p\to 3\pi^0 p\to 6\gamma p$ combined with
 $\gamma p\to\pi^0\eta p\to4\pi^0 p\to 8\gamma p$.
 As seen, in the six-photon data sample, the background level under the $\eta'$ peak
 is significantly larger than the $\eta'$ signal.
 The events reconstructed by testing the $\gamma p\to \pi^0\pi^0\eta p\to 6\gamma p$ hypothesis (with the three
 additional invariant-mass constraints) and selected with the corresponding CL$>$1\%
 are shown in Fig.~\ref{fig:etapr_eta2pi0_ept2014_4x1}(b).
 The background level is much smaller here, but it is not
 sufficient yet for reliable analysis of $\eta'\to \pi^0\pi^0\eta\to 6\gamma$ decays.
 The result of suppressing background contributions with $\CL(\gamma p\to \pi^0\pi^0\eta p\to 6\gamma p)>0.04$
 for the main hypothesis and $\CL(\gamma p\to\pi^{0}\pi^{0}\pi^{0}p)<0.0075$ for the background hypothesis
 is shown in Fig.~\ref{fig:etapr_eta2pi0_ept2014_4x1}(c).
 As seen, the background level under the $\eta'$ peak became sufficiently small not to cause much impact
 on the analysis of $\eta'\to \pi^0\pi^0\eta$ decays.
 However, the background suppression resulted also in discarding
 20\% of good $\eta'\to \pi^0\pi^0\eta \to 6\gamma$ decays, preventing from further
 tightening background cuts. The same figure also includes the invariant-mass
 distributions for the events that did not pass the hypothesis also involving the fourth additional constraint
 on the $\eta'$ mass and with the corresponding CL$(\gamma p\to \eta' p\to \pi^0\pi^0\eta p\to 6\gamma p)<4\%$ for them.
 Because of such a cut, the latter distributions have a dip in the region of the $\eta'$ mass.  
 Then the events from Fig.~\ref{fig:etapr_eta2pi0_ept2014_4x1}(c) for which
 CL$(\gamma p\to \eta' p\to \pi^0\pi^0\eta p\to 6\gamma p)>4\%$ are shown
 in Fig.~\ref{fig:etapr_eta2pi0_ept2014_4x1}(d), illustrating good agreement in the energy calibration
 and resolution for the experimental and MC $\eta'\to \pi^0\pi^0\eta\to 6\gamma$ decays.
 The events shown in Fig.~\ref{fig:etapr_eta2pi0_ept2014_4x1}(d) also represent
 the final experimental and MC samples of $\eta'\to \pi^0\pi^0\eta\to 6\gamma$ decays, but with the kinematics
 reconstructed from the results of testing the $\gamma p\to \eta' p\to \pi^0\pi^0\eta p\to 6\gamma p$
 hypothesis, involving the fourth additional constraint on the $\eta'$ mass.
 The estimation of background remaining in the final experimental sample is demonstrated
 in Fig.~\ref{fig:etapr_eta2pi0_ept2014_4x1}(d) by the solid magenta line.
 The normalization of this background is based on the spectra
 shown in Fig.~\ref{fig:etapr_eta2pi0_ept2014_4x1}(c), where the combined MC simulation of the
 background reactions is normalized on the parts of the experimental spectrum that are away from
 the $\eta'$ signal. For the case demonstrated in Fig.~\ref{fig:etapr_eta2pi0_ept2014_4x1}(d),
 the fraction of the remaining background is estimated as 5.5\%.
 With a more conservative background normalization in Fig.~\ref{fig:etapr_eta2pi0_ept2014_4x1}(c),
 the background fraction in the final sample could be larger, reaching up to 7.5\%.
 
 Based on the MC simulation, the background events were found to be distributed randomly over
 the Dalitz plot, looking similar to the distribution from the phase-space MC simulation
 of $\eta'\to \pi^0\pi^0\eta\to 6\gamma$ decays.
 Thus, such a behavior of the background events cannot result in mimicking any narrow structure as cusp,
 but the Dalitz-plot slopes, as well as a cusp structure, could look more shallow, introducing corresponding
 systematic effects in the extracted matrix-element parameters.
 The acceptance-corrected $\eta'\to \pi^0\pi^0\eta$ Dalitz plot from
 Analysis II is shown in Fig.~\ref{fig:PatrikDP}(b). Compared to Analysis I, the latter plot has smaller binning
 and includes bins that also cover the physical boundaries of the $\eta'\to \pi^0\pi^0\eta$ Dalitz plot.
 To include the boundary bins in a fit, the density function should be corrected for phase space
 available in those bins.
\begin{figure*}
\includegraphics[width=0.95\textwidth]{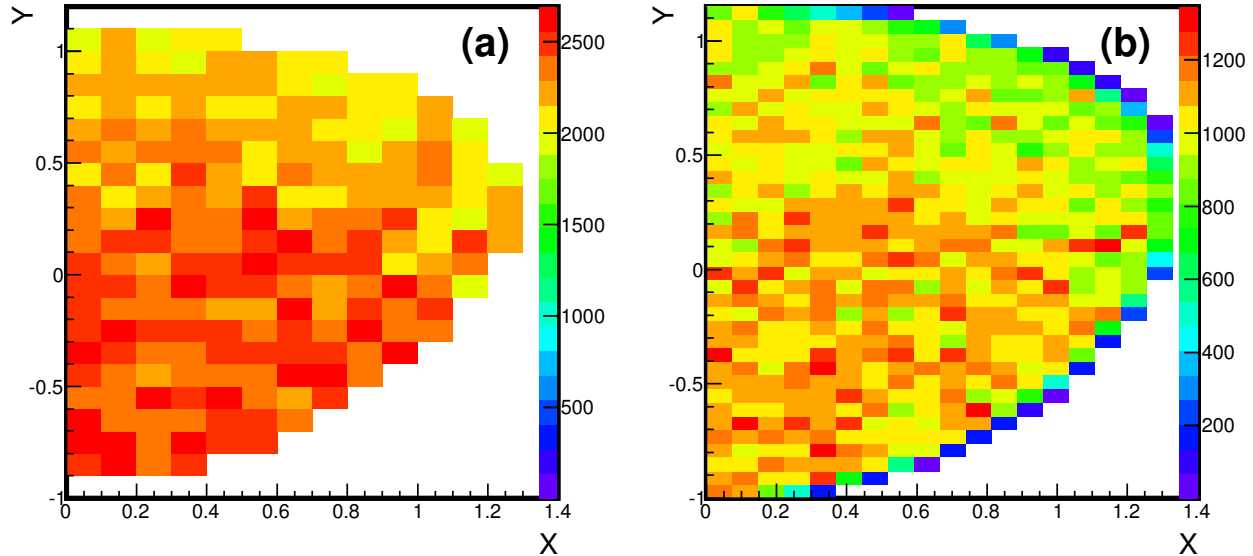}
\caption{ Experimental acceptance-corrected $\eta'\to \pi^0\pi^0\eta$ Dalitz plot obtained from (a) Analysis I
 and (b) Analysis II. See text for details of the two analyses.}
\label{fig:PatrikDP}
\end{figure*}

 Other informative distributions that illustrate the deviation of the actual $\eta'\to \pi^0\pi^0\eta$ decay
 from phase space are ratios of the experimental $X$, $Y$, $m(\pi^0\pi^0)$, and $m(\pi^0\eta)$ spectra
 to their phase-space MC simulation.
 In Analysis I, those background-free experimental spectra were obtained with a procedure similar to that used
 to measure the Dalitz plot itself and with the same selection criteria.
 The ratios obtained for each observable are shown in Fig.~\ref{fig:DatadivPS}
 by blue open squares. The results for larger masses in the $m(\pi^0\pi^0)$ and $m(\pi^0\eta)$ spectra are
 not included in Fig.~\ref{fig:DatadivPS}, as the fitting procedure used to measure the experimental signal
 in those bins was giving large uncertainties in the results. 
 The ratios from Analysis II are depicted in Fig.~\ref{fig:DatadivPS} by red open circles. Because the 
 experimental sample includes the remaining background events, the systematic uncertainties that reflect the level of this background were added linearly to the statistical uncertainties. The magnitudes of these
 systematic uncertainties were determined from the change in the ratios depending on the kinematic-fit $\CL$
 used for the event selection, under the assumption that such a change occurs solely due to a different level
 of the remaining background. The MC simulation of the two background reactions demonstrated that the spectra
 with the remaining background are close to the $\eta'\to \pi^0\pi^0\eta$ phase-space behavior and cannot
 mimic any narrow structures when divided by the corresponding phase-space spectra from the $\eta'$ MC simulation. 
 The normalization of the ratio distributions is based on the ratio in the number of events in the experimental
 and MC spectra. A smaller binning in $Y$ and $m(\pi^0\pi^0)$ was chosen such that the expected cusp
 structure could be visible. Figure~\ref{fig:DatadivPS} shows that the data points from Analyses I and II are
 in good agreement within their uncertainties, but demonstrate a significant deviation from phase space.
 These points could be used, together with
 the Dalitz plots, for testing different models; therefore, all data points from the Dalitz plots
 and the four ratio distributions are provided as supplemental material to this paper.

\section{Results and Discussion}
\label{sec:Results}

 In Analysis I, to determine the $\eta'\to \pi^0\pi^0\eta$ matrix-element parameters, the Dalitz-plot fitting
 procedure was based on the minimization of
\begin{equation}
\chi^2 = \sum_{i=0}^{n_{\rm bins} } \left(\frac{N_{\rm{exp}}^{i}- \epsilon_{i} \cdot f_i(X, Y)}{\sigma_{i}} \right)^2~,
\end{equation}  
 where, for bin $i$, $N_{\rm exp}^{i}$ is the measured number of $\eta'$ mesons, $\epsilon_{i}$ is
 the corresponding detection efficiency, and $f_i(X, Y) \sim |M_i(X,Y)|^2$ is the theoretical function used in the fit and
 integrated over the given bin. The uncertainty $\sigma_i$ includes both the uncertainties in $N_{\rm exp}^{i}$
 and in $\epsilon_{i}$.

 In Analysis II, the $\chi^2$ calculation was based on the differences between the bin contents of the
 measured (i.e., uncorrected for the acceptance) Dalitz plot and the corresponding plot with
 the $\eta'\to \pi^0\pi^0\eta$ MC events weighted with the fit function.
 For every fit iteration, a weight of each MC event was calculated
 with the fit function taken from the generated values for $X$ and $Y$ and with its current parameters
 taken from the fit. Then the entry in the MC Dalitz plot was based on the reconstructed
 $X$ and $Y$, thus taking the experimental acceptance and resolution into account.
\begin{figure*}
\includegraphics[width=1.0\textwidth]{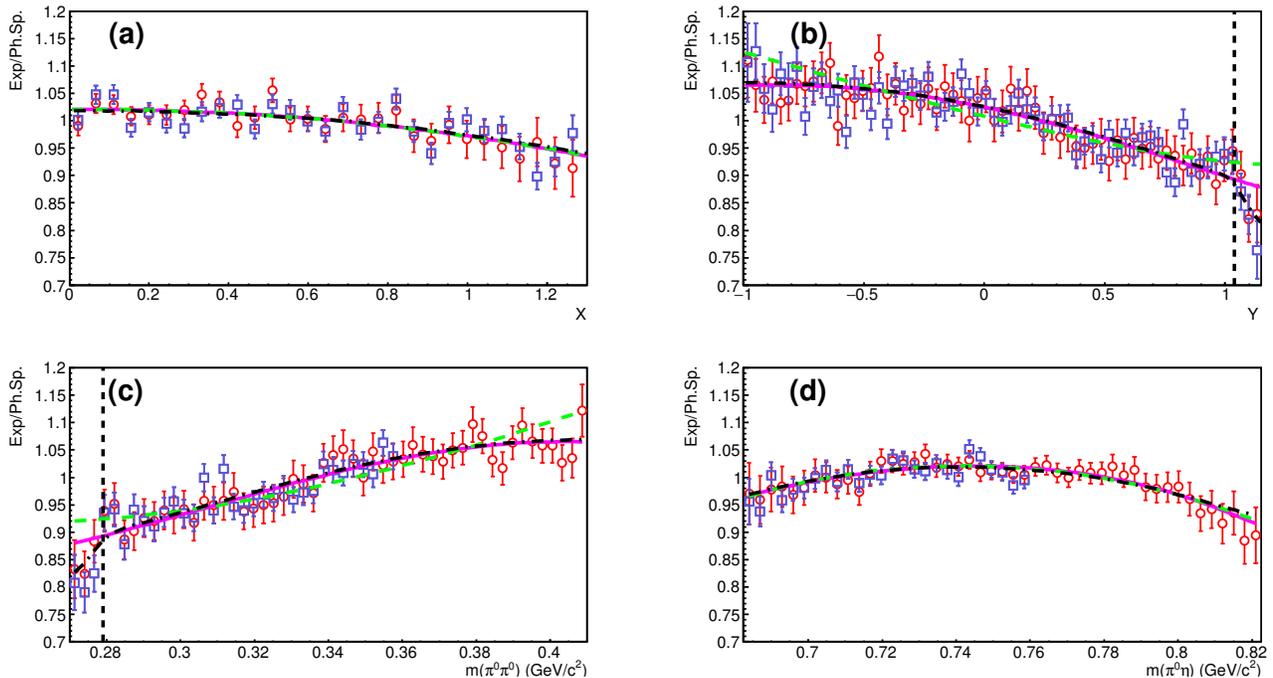}
\caption{Ratios of the $\eta'\to \pi^0\pi^0\eta$ experimental distributions for (a) $X$, (b) $Y$, (c) $m(\pi^0\pi^0)$,
 and (d) $m(\pi^0\eta)$ to their phase-space MC simulation normalized to the
 experimental number of events. The data points from Analysis I are shown by open blue squares and from Analysis II
 by open red circles. The vertical dashed lines in (b) and (c) show the position corresponding to the mass of
 two charged pions. The fit results for the $\eta'\to \pi^0\pi^0\eta$ Dalitz plot from Analysis I
 with Eq.~(\protect\ref{eqn:M1sq}) are shown by the magenta solid lines, with Eq.~(\protect\ref{eqn:M2sq})
 by the green dashed lines, and with the NREFT amplitude~\protect\cite{Kubis:2009sb} by the black dash-dotted lines.}
\label{fig:DatadivPS}
\end{figure*} 

 The main results from fitting experimental Dalitz plots of Analyses I and II with different functions are listed
 in Table~\ref{tab:tableDP}. The first uncertainty in parameter values represents the errors from the fits.
 The systematic uncertainties in the matrix-element parameters were evaluated only for its standard parametrization
 with Eq.~(\ref{eqn:M1sq}). The study of systematic effects was more scrupulous in Analysis I, which provides 
 a near background-free Dalitz plot, but whose results could be sensitive to the fitting procedure of measuring
 the $\eta'$ signal in every bin and to the experimental resolution, which smeared $\eta'$ events out of the physical region. To estimate systematic uncertainties (given as the second uncertainty) in the matrix-element parameters,
 their sensitivity was tested to the changes in the procedure measuring the signal, the $\CL$ selection criteria,
 the Dalitz-plot bin width, and the period of data taking.
\begin{table*}[t]
\centering
\begin{tabular}{|c|l|c|c|c|c|c|}
\hline
 Fit \# & $|M(X,Y)|^2 \sim 1+aY+bY^2+dX^2$ & $\chi^2/$dof & $a$ & $b$ & $d$ &  \\ \hline  \hline
 1 & Analysis I & 1.092 & $-0.074(8)(6)$ & $-0.063(14)(5)$ & $-0.050(9)(5)$ &  \\ \hline
 2 & Analysis II & 1.100 & $-0.071(7)(2)$ & $-0.069(11)(4)$ & $-0.060(8)(5)$ &  \\ \hline  \hline
   & $|M(X,Y)|^2 \sim 1+aY+bY^2+dX^2+\kappa_{21}YX^2$ &  & $a$ & $b$ & $d$ & $\kappa_{21}$ \\ \hline  \hline
 3 & Analysis I & 1.092 & $-0.069(10)$ & $-0.060(14)$ & $-0.043(11)$ & $-0.026(24)$ \\ \hline  
 4 & Analysis II & 1.096 & $-0.062(8)$ & $-0.066(12)$ & $-0.052(9)$ & $-0.034(19)$ \\ \hline \hline 
   & $|M(X,Y)|^2 \sim |1+\alpha Y|^2+dX^2$ & & Re($\alpha$) & Im($\alpha$) & $d$ & \\ \hline \hline
 5 & Analysis I & 1.197 & $-0.047(4)$ & $0.000(40)$ & $-0.037(9)$ &  \\   \hline
 6 & Analysis II & 1.171 & $-0.045(3)$ & $0.000(31)$ & $-0.046(8)$ &  \\   \hline \hline
  & NREFT amplitude &  & $a'$ & $b'$ & $d'$ & $a_0-a_2$ \\ \hline \hline
 7 & Analysis I & 1.094 & $-0.155(8)$ & $-0.026(20)$ & $-0.048(10)$ & $0.2644$(fix) \\   \hline 
 8 & Analysis I & 1.095 & $-0.149(6)$ & $-0.026(8)$ & $-0.048(7)$ & $0.191(79)$ \\   \hline 
 9 & Analysis I & 1.091 & $-0.155$(fix)& $-0.026$(fix) & $-0.048$(fix) & $0.255(90)$ \\   \hline 
 10 & Analysis II & 1.092 & $-0.142(7)$ & $-0.035(12)$ & $-0.063(8)$ & $0.2644$(fix) \\   \hline 
 11 & Analysis II & 1.088 & $-0.142$(fix) & $-0.035$(fix) & $-0.063$(fix) & $0.262(58)$ \\   \hline 
\end{tabular}
\caption{ Results of fitting the $\eta'\to \pi^0\pi^0\eta$ Dalitz plots with different functions describing
 its density. The first uncertainty in parameter values represents their errors
 from the fits. The second error, which represents the systematic uncertainty, is evaluated for the fit
 with Eq.~(\protect\ref{eqn:M1sq}) only. The parameters held fixed during the fits with
 the NREFT amplitude~\protect\cite{Kubis:2009sb} are marked as (fix).} 
\label{tab:tableDP}
\end{table*}

 Cross checks of systematic effects in Analysis I, which are divided into two categories, are summarized in Table~\ref{tab:syst}.
 All tests that were made without changing the selection
criteria for events in the final data sample were included into the first category.
 Test \#1 accumulates the tests made to check the procedure that measures the $\eta'$ signal. Those tests
 included changes in the order of the polynomial used to describe the background in each Dalitz-plot bin, by varying
 it from 2 to 5. Also, instead of using the signal line shape from its MC simulation, the number of $\eta'$ decays was
 obtained by subtracting the background from
 the experimental spectra, based on the fit results for polynomial of order 4, describing the background contribution.
 In test \#2, the bin width in the m($\pi^0\pi^0\eta$) spectra, used to measure the $\eta'$ signal, varied from 2 to 8~MeV/$c^2$.
 In test \#3, the Dalitz-plot bin width varied from 0.05 to 0.15 in both $X$ and $Y$.
 Because none of these tests reveals a deviation in parameter values more than the corresponding fit errors,
 they did not contribute to the systematic uncertainties of the matrix-element parameters.
 
 The second category included all tests that were made by changing experimental statistics.
 The consistency of the results was checked by comparing the uncorrelated differences~\cite{stat13} between
 the parameter values obtained from fitting to the final, $x_f\pm \sigma_f$, and test, $x_t\pm \sigma_t$, Dalitz plots:
\begin{equation}
\Delta x_{\rm uncor} = \frac{|x_{f}-x_{t}|}{\sqrt{|\sigma^2_f-\sigma^2_t}|}~, 
\label{eqn:statuncfit}
\end{equation}
 with a systematic effect revealed if $\Delta x_{\rm uncor}>2$.

 Because the total data set was collected during three different periods of data taking,
 the consistency of these three subsets with each other was checked for the Dalitz plots with the main bin width,
 which was 0.1 in both $X$ and $Y$ (test \#4), and with the bins enlarged to 0.15 (test \#5), which decreased
 the corresponding statistical uncertainties closer to the level of the final Dalitz plot. 
 The largest deviation, $\Delta x_{\rm uncor}=2.0$, was found for parameter $a$ with the bin width 0.1, but it became
 smaller with the enlarged bin width. Based on the results of these tests, it was concluded that the data from
 the different periods of data taking are consistent with each other, and there are no systematic differences
 between them.

 Test \#6 involved changes in the selection criteria based on the kinematic-fit $\CL$ for
 the $\gamma p\to \pi^0\pi^0\eta p\to 6\gamma p$ and $\gamma p\to \pi^0\pi^0\pi^0 p\to 6\gamma p$ hypotheses.
 Variation of $\CL$ values results in both a different level of background events in the experimental
 $m(\pi^0\pi^0\eta)$ spectra and a change in the resolution of selected experimental and MC events.
 As Table~\ref{tab:syst} shows, the value of $\Delta x_{\rm uncor}$ for each parameter exceeded the magnitude of 2,
 which was chosen to expose a systematic effect. The corresponding systematic uncertainties were then
 taken as half of the difference between the maximum and minimum parameter values obtained in test \#6.
Such an evaluation could be quite conservative as the tests giving parameters
between their maximum and minimum values, as well as the parameter errors
in individual fits, are neglected in this evaluation.
 Taking these systematic uncertainties into account resulted in the following values for the $\eta'\to \pi^0\pi^0\eta$
 matrix-element parameters:  
\begin{equation}
\begin{aligned}
a &= -0.074(8)_{\rm{stat}}(6)_{\rm{syst}}, \\ 
b &= -0.063(14)_{\rm{stat}}(5)_{\rm{syst},}  \\
d &= -0.050(9)_{\rm{stat}}(5)_{\rm{syst}}.
\label{eqn:DPfinalresults}
\end{aligned}
\end{equation}
 For convenience, these results are also plotted in Fig.~\ref{fig:DPsummary}, which compares the existing
 results and calculations for the standard matrix-element parametrization. For similarity with previous measurements,
 the correlation matrix is provided for fit \#1, the results of which were taken as the main results
in (8):
 
 \begin{equation}
 \begin{pmatrix}
 	\begin{matrix}
		 & b & d \\
		 a  & -0.542 & -0.289  \\
		 b  & & 0.294 \\
 	\end{matrix}
 \end{pmatrix}
 \end{equation}

\begin{table*}[t]
\centering
\begin{tabular}{|l||l|l|l|l|l|l|l|}
\hline
 Test \# & Category 1 & $a_{\rm min,max}$ & $|\Delta a|/ \sigma_a$ & $b_{\rm min,max}$ & $|\Delta b|/ \sigma_b$ & $d_{\rm min,max}$ & $|\Delta d|/ \sigma_d$ \\ \hline  \hline
1 & Signal estimation &  $^{-0.075(8)}_{-0.072(8)}$ & 0.1 & $^{-0.073(14)}_{-0.061(14)}$ & 0.8& $^{-0.050(9)}_{-0.048(9)}$ &0.2\\ \hline  

 2 &  $m(\pi^0\pi^0\eta)$ bin width & $^{-0.081(8)}_{-0.074(8)}$ & 0.8 & $^{-0.069(14)}_{-0.063(14)}$ & 0.2 & $^{-0.051(9)}_{-0.048(9)}$& 0.1\\ \hline  
 3 &  Dalitz-plot bin width & $^{-0.082(8)}_{-0.068(8)}$ & 1.0 & $^{-0.067(15)}_{-0.052(13)}$ & 0.8 & $^{-0.050(10)}_{-0.045(9)}$ & 0.6\\ \hline \hline
 & Category 2  & $a_{\rm min,max}$ & $\Delta a_{\rm uncor}$ & $b_{\rm min,max}$ & $\Delta b_{\rm uncor}$ & $d_{\rm min,max}$ & $\Delta d_{\rm uncor}$  \\ \hline  \hline
 4 & 3 periods, bin width 0.1 & $^{-0.097(14)}_{-0.064(14)}$ & 2.0 &  $^{-0.085(23)}_{-0.055(25)}$ & 1.2 & $^{-0.055(15)}_{-0.042(16)}$ & 0.6\\ \hline 
 5 & 3 periods, bin width 0.15& $^{-0.087(14)}_{-0.057(15)}$ & 1.7 &  $^{-0.086(25)}_{-0.054(27)}$ & 0.9 & $^{-0.061(17)}_{-0.046(18)}$ & 0.8\\ \hline 
 6 & $\CL$ cuts	& $^{-0.084(9)}_{-0.071(8)}$ & 3.9 &  $^{-0.070(14)}_{-0.061(15)}$ & 2.4 & $^{-0.057(10)}_{-0.046(9)}$ & $2.3$\\ \hline 
\end{tabular}
\caption{ Cross checks of systematic effects in the matrix-element parameters obtained for Eq.~(\ref{eqn:M1sq}).}
\label{tab:syst}
\end{table*}

 In Analysis II, it was found that the systematic uncertainties in the parameter values are mostly
 caused by the background remaining in the selected $\eta'\to \pi^0\pi^0\eta$ experimental decays.
 Similar to Analysis I, their magnitudes were determined by comparing the results from fitting the experimental
 Dalitz plots that were obtained with looser and tighter selection criteria on the kinematic-fit $\CL$
 for both the $\gamma p\to \eta' p\to \pi^0\pi^0\eta p\to 6\gamma$ and 
 $\gamma p\to \pi^0\pi^0\pi^0 p\to 6\gamma p$ hypotheses. This assumes that the changes
 in the parameter values are solely caused by different fractions of the remaining background, rather than
 rejecting $\eta'$ decays with poorer resolution by tightening
 the $\CL(\gamma p\to \eta' p\to \pi^0\pi^0\eta p\to 6\gamma)$ criterion, in combination with
 lowering experimental statistics. The results of Analysis II for the standard parametrization of
 the $\eta'\to \pi^0\pi^0\eta$ Dalitz plot are listed in Table~\ref{tab:tableDP} (\#2), demonstrating
 good agreement with Analysis I (\#1) within the uncertainties.
 Similarly to the results for the $X$, $Y$, $m(\pi^0\pi^0)$, and $m(\pi^0\eta)$ spectra
 of Analysis II, the systematic uncertainties in the results for the standard parameters
 should be added linearly to the fit uncertainties.
 Because the results of Analysis II were obtained by fitting the
measured plot, it was also checked that fitting the corresponding
acceptance-corrected Dalitz plot, shown in Fig.~6(b), gives similar
results. For the standard matrix-element parametrization,
such a fit resulted in $a=-0.071(7)$, $b=-0.069(11)$, and $d=-0.061(7)$, with $\chi^2/{\rm dof}=1.085$,
 demonstrating good agreement with fit \#2 and, in such way, confirming the reliability
 of the procedure used for the acceptance correction, which minimizes the smearing effect from the
 experimental resolution. For comparison, the fit to the acceptance-corrected Dalitz plot obtained
 with just a phase-space MC simulation results in 
 $a=-0.073(7)$, $b=-0.061(11)$, and $d=-0.056(8)$, with $\chi^2/{\rm dof}=1.123$, which deviates
 more from fit \#2, but still is in good agreement with it and fit \#1 within the uncertainties.  

 The present results of Analysis I (shown also in Fig.~\ref{fig:DPsummary}) and II are consistent with the previous
 $\eta'\to\pi^0\pi^0\eta$ measurement by GAMS-4$\pi$, $a=-0.067(16)(4)$, $b=-0.064(29)(5)$,
 and $d=-0.067(20)(3)$~\cite{Blik:2009zz}, but improve upon their uncertainties.
 The agreement with the $\eta'\to\pi^+\pi^-\eta$ results is better with the BESIII~\cite{Ablikim:2010kp}, compared to the VES data~\cite{Dorofeev:2006fb}.
 At the same time, the present results do not improve much the situation existing between the experimental data and
 the calculations~\cite{Borasoy:2006uv,Escribano:2010wt,Fariborz:2014gsa}. For instance, the results obtained for parameter $a$ deviate from
the prediction of the $U(3)$ chiral effective-field theory~\cite{Borasoy:2006uv},
but are consistent with the value from the generalized linear-sigma model~\cite{Fariborz:2014gsa}.
 The situation is opposite for parameters $b$ and $d$, for which
only the latter model cannot reproduce the experimental results.
It is expected, however, that the agreement with the linear-sigma model
could be improved if higher-order corrections were included in the
calculations~\cite{Fariborz_privat}.
 
 The quality of fitting the experimental Dalitz plots with its standard parametrization, based on Eq.~(\ref{eqn:M1sq}),
 can be seen from both the $\chi^2/$degrees-of-freedom ($\chi^2/$dof) values listed in Table~\ref{tab:tableDP}, and
 the fit results compared to the $X$, $Y$, $m(\pi^0\pi^0)$, and $m(\pi^0\eta)$ spectra shown in Fig.~\ref{fig:DatadivPS}. 
 As seen in this figure, the Dalitz-plot fit results, depicted by the magenta solid line for Eq.~(\ref{eqn:M1sq}), are
 in good agreement with the experimental data points, except for the region in the $Y$ and $m(\pi^0\pi^0)$ spectra
 where the cusp structure was expected.

 More fits were made to check the sensitivity of the measured Dalitz plots to the higher-order terms $\kappa_{21}YX^2$
 and $\kappa_{40}X^4$, the magnitudes of which were evaluated in Ref.~\cite{Escribano:2010wt}.
 The fit results after adding only the $YX^2$ term are listed in Table~\ref{tab:tableDP} (\#3, 4).
 They show practically no improvement in $\chi^2/$dof. 
 The magnitude found for parameter $\kappa_{21}$ is somewhat larger
 than values predicted in Ref.~\cite{Escribano:2010wt} (with the closest prediction $\kappa_{21}=-0.009(2)$),
 but its large uncertainty and correlation with other parameters
 cannot justify the results obtained. A similar situation was observed after adding the $\kappa_{40}X^4$ term, which resulted in no
 improvement in $\chi^2/$dof and which had a strong correlation with the $\kappa_{21}X^2$ term.   
 It appears that $\eta'\to \pi\pi\eta$ data with much higher statistical accuracy are needed for more reliable
 estimation of possible $YX^2$ and $X^4$ contributions. 

As discussed in Section~\ref{Intro}, a negative value of parameter $b$ excludes a linear $Y$ dependence of the $\eta'\to \pi\pi\eta$ decay amplitude, with the matrix element parametrized according to Eq.~(\ref{eqn:M2sq}). To allow a comparison to earlier measurements~\cite{Briere:1999bp,Blik:2009zz,Ablikim:2010kp}, this parametrization was also tested with the present data. The numerical results for Analyses I and II are listed in Table~\ref{tab:tableDP} (\#5, 6), and the fit from Analysis I is shown in Fig.~\ref{fig:DatadivPS} by the green dashed lines. As seen, the $\chi^2/$dof value becomes worse compared to the standard parametrization and both the $Y$ distribution and the related $m(\pi^0\pi^0)$
 spectrum cannot be described with such a linear dependence. Nevertheless, both Analyses I and II give very similar results
 for Re($\alpha$) and Im($\alpha$), which are also consistent with the most recent $\eta'\to \pi^0\pi^0\eta$ results
 from GAMS-4$\pi$: Re($\alpha$) = $-$0.042(8) and Im($\alpha$) = 0.00(7)~\cite{Blik:2009zz}.

 As also seen in Fig.~\ref{fig:DatadivPS}, both Analyses I and II indicate a cusp at the $\pi^+\pi^-$ threshold in the $Y$ and $m(\pi^0\pi^0)$ spectra (marked by the vertical dashed lines),
 and none of the parametrizations tested so far were able to describe this region.
 To investigate this effect, the experimental Dalitz plots were fitted with the $\eta'\to \pi\pi\eta$ decay
 amplitude parametrized within the NREFT framework~\cite{Kubis:2009sb}.
 In this model, the decay amplitude is decomposed up to two loops, 
$A(\eta'\to \pi\pi\eta) = A^{\rm tree} + A^{\rm 1-loop} + A^{\rm 2-loop}$,
with the tree amplitude complemented by final-state interactions of one and two loops.
The one- and two-loop amplitudes that describe the final-state interactions are calculated in
 Ref.~\cite{Kubis:2009sb} based on the magnitudes of $\pi\pi$ scattering lengths, which were previously extracted
 from the analysis of $K \to 3\pi$ decays~\cite{Colangelo:2006va, Bissegger:2007yq, Bissegger:2008ff}.
 In the same work, there are no calculations for the tree amplitude.
 To obtain couplings for the $\eta'\to \pi\pi\eta$ Lagrangian by matching the standard Dalitz-plot parametrization,
 $|M(X,Y)|^2 \sim 1 + a'Y + b'Y^2 + d'X^2 +...$, the tree amplitude can be parametrized as
\begin{equation}
 A^{\rm tree}(X,Y) \sim 1 + \frac{a'}{2}Y + \frac{1}{2}(b'-\frac{a'^2}{4})Y^2 + \frac{d'}{2}X^2 +...~,
\label{eqn:atree_nreft}
\end{equation}
 where $a'$, $b'$, and $d'$ are the tree-amplitude parameters that describe the $\eta'\to \pi\pi\eta$ dynamics
 before the contributions from final-state interactions. Those parameters are also involved in the calculation
 of the one- and two-loop amplitudes. Then the total amplitude determines the final dependence on $Y$ and $m(\pi^0\pi^0)$.
 At the same time, the dependence on $X$ and $m(\pi^0\eta)$ is still mostly defined by the tree amplitude
 and its parameter $d'$, as $\pi\eta$ final-state interactions turned out to be very small and were neglected
 in the NREFT approach.
 Because the tree amplitude is the same for both the $\eta'\to \pi^0\pi^0\eta$ and $\eta'\to \pi^+\pi^-\eta$ decays,
 the difference in them is determined solely by the final-state interactions, which also produce the cusp structure
 in the spectra from the neutral decay mode. The magnitude and the sign of this cusp structure is mostly determined by
 the scattering length combination $a_0-a_2=0.2644\pm 0.0051$, where $a_0=0.220\pm 0.005$ and
 $a_2=-0.0444\pm 0.0010$~\cite{Colangelo:2001df, Kubis:2009sb}.
  
 Because, in Analysis I with the Dalitz-plot bin width 0.1, the cusp region was mostly located in the boundary bins
 that were rejected from the fits, the experimental plot was remeasured with a narrower bin width, 0.05,
 in both $X$ and $Y$, which allowed a better access to the cusp.
 The results of fitting this Dalitz plot with the NREFT amplitude, having only three free parameters from
 the tree term, are listed in Table~\ref{tab:tableDP} (\#7), and the fit is shown in Fig.~\ref{fig:DatadivPS}
 by the black dash-dotted lines.
 The results obtained here for the tree-amplitude parameters are strongly
 model dependent and can be used only for qualitative purposes. 
 From the comparison with the fit results based on the standard parametrization (\#1),
 the $\chi^2/$dof value is of the same magnitude, but the NREFT fit describes better the $Y$ and $m(\pi^0\pi^0)$ data points below the $\pi^+\pi^-$ threshold. The magnitude of parameters $a'$ and $b'$, reflecting the decay dynamics
 before final-state interaction, look now closer to the linear $Y$ dependence,
 which is typically expected for the tree amplitude~\cite{Alde:1986nw, Escribano:2010wt}.
 As expected, the magnitude of parameter $d'$ is very close to $d$ from the fit with the standard parametrization. The parameters obtained for the tree amplitude in the fits with
the NREFT amplitude (\#7-\#11 in Table I) could be compared
to the corresponding calculations for the leading-order terms
of the $\eta'\to \pi^0\pi^0\eta$ decay amplitude, which are, e.g.,
provided in Ref.~\cite{Escribano:2010wt}.

To test more reliably if the structure seen below the $\pi^+\pi^-$ threshold
is in agreement with the magnitude of the cusp predicted by NREFT,
the difference $a_0-a_2$ was implemented in the NREFT code as its
free parameter,
and the fit repeated with the four parameters. The results of this fit
are listed in Table~\ref{tab:tableDP} (\#8). As seen, the four-parameter fit does not improve the $\chi^2/$dof value.
  Furthermore, such a fit is not well justified as the overall $Y$ and
 $m(\pi^0\pi^0)$ distributions in NREFT depend on the individual values
of $a_0$ and $a_2$ as well.
 A fairer procedure would be to fix the three tree-amplitude parameters according to the fit
 made with fixed $a_0-a_2$ (\#7), and then to release only this difference to test solely the cusp region.
 The results for the latter fit are listed in Table~\ref{tab:tableDP} (\#9). It demonstrates good consistency
 with the known value $a_0-a_2=0.2644\pm 0.0051$~\cite{Colangelo:2001df}, though with much larger uncertainties, compared to it.

 The results of fitting the $\eta'\to \pi^0\pi^0\eta$ Dalitz plot from Analysis II with the NREFT amplitude,
 which are also listed in Table~\ref{tab:tableDP} (\#10, 11), confirm the corresponding results obtained from Analysis I.
 All fit results obtained with the NREFT amplitude speak for the quality of this model, together with the previously extracted scattering-length combinations.
   
 \begin{figure*}
\includegraphics[width=0.95\textwidth]{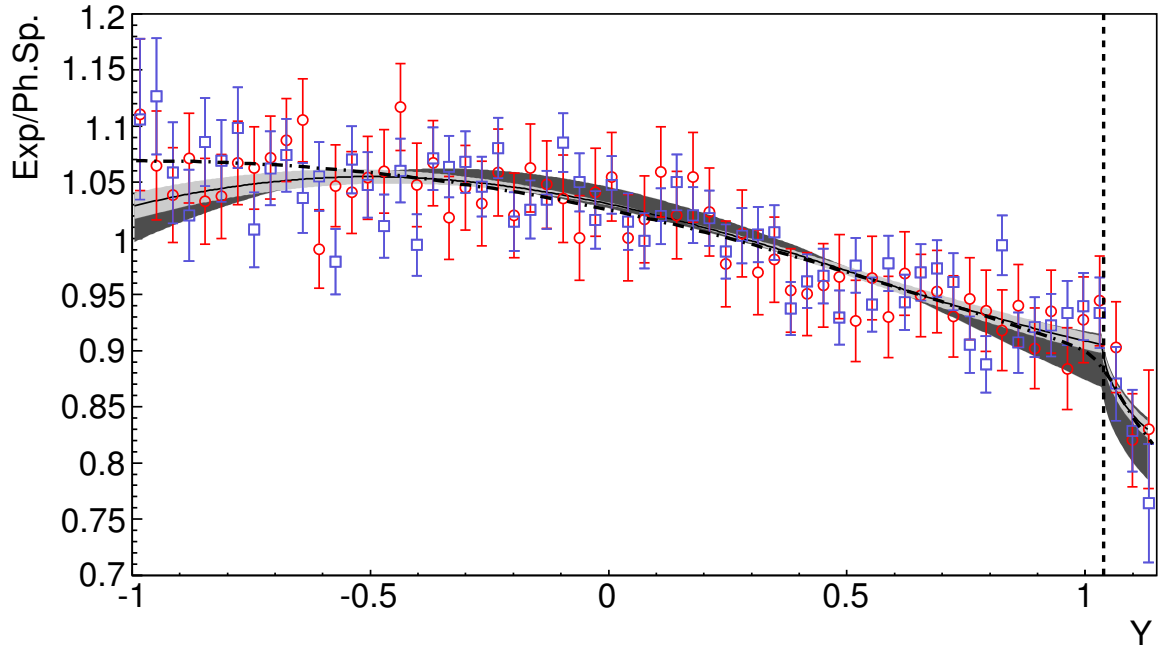}
\caption{ Same as Fig.~\ref{fig:DatadivPS}(b), including the NREFT fit \#7, but compared to the dispersive analysis
 of the $\eta'\to\pi\pi\eta$ decay amplitude from Ref.~\protect\cite{Isken:2017dkw}, the prediction
 of which is shown for the case of fitting to the BESIII data~\protect\cite{Ablikim:2010kp} with three subtraction
 constants. The light-gray error band represents the uncertainties obtained from fitting to the data,
 and the dark-gray band from the variation of the phase input.}
\label{fig:DatadivPSKubisIsken}
\end{figure*} 

 In Fig.~\ref{fig:DatadivPSKubisIsken}, the data points for the $Y$ dependence and the corresponding NREFT fit \#7
 are also compared to the dispersive analysis
 of the $\eta'\to\pi\pi\eta$ decay amplitude from Ref.~\cite{Isken:2017dkw}. Namely, their prediction based on
 fitting the BESIII data~\cite{Ablikim:2010kp} with three subtraction constants is shown, including
 two error bands representing different uncertainties: one from fitting to the data and the other from 
 the variation of the phase input. As the figure shows, the present data points are in good agreement within the uncertainties
 with this most recent calculation.

\section{Summary and conclusions}
\label{sec:Conclusion}
An experimental study of the $\eta'\to \pi^0\pi^0\eta \to 6\gamma$ decay
has been conducted with the best up-to-date statistical accuracy, by
measuring $\eta'$ mesons produced in the $\gamma p \to \eta' p$ reaction
with the A2 tagged-photon facility at the Mainz Microtron, MAMI.
The results of this work obtained for the standard parametrization of the $\eta'\to \pi^0\pi^0\eta$
matrix element, $a = -0.074(8)_{\rm{stat}}(6)_{\rm{syst}}$, $b = -0.063(14)_{\rm{stat}}(5)_{\rm{syst}}$ and $d = -0.050(9)_{\rm{stat}}(5)_{\rm{syst}}$ are consistent with the most recent results for $\eta'\to\pi\pi\eta$ decays,
but have smaller uncertainties. It was tested that including higher-order terms
does not improve the description of the $\eta'\to \pi^0\pi^0\eta$ Dalitz plot.
The available statistics and experimental resolution allowed,
for the first time, an observation of a structure below the
$\pi^+\pi^-$ mass threshold, the magnitude and sign of which,
checked within the framework of the nonrelativistic effective-field theory,
demonstrated good agreement (within a one-$\sigma$ level) with the
cusp that was predicted
based on the $\pi\pi$ scattering length combination, $a_0-a_2$,
extracted from $K \to 3\pi$ decays. The data points from the experimental Dalitz plots
and ratios of the $X$, $Y$, $m(\pi^0\pi^0)$, and $m(\pi^0\eta)$ distributions to phase space
are provided as supplemental material to the paper.

\begin{acknowledgments}
The authors wish to acknowledge the excellent
support of the accelerator group and operators of MAMI.
We would like to thank Bastian Kubis for providing
the code of NREFT and to thank him, Tobias Isken, Peter Stoffer, Pere Masjuan, and Stefan Leupold for fruitful discussions of our results. 
This work was supported by the Deutsche Forschungsgemeinschaft
(SFB443, SFB/TR16, and SFB1044, PRISMA Cluster of Excellence), the European Community-Research
Infrastructure Activity under the FP6 ``Structuring the European Research Area"
programme (Hadron Physics, Contract No. RII3-CT-2004-506078),
Schweizerischer Nationalfonds (Contract Nos. 200020-156983,
132799, 121781, 117601, 113511), the UK Science and
Technology Facilities Council (STFC 57071/1, 50727/1),
the U.S. Department of Energy (Offices of Science and Nuclear Physics,
 Award Nos. DE-FG02-99-ER41110, DE-FG02-88ER40415, DE-FG02-01-ER41194, DE-SC0014323), National Science Foundation (Grant Nos. PHY-1039130, IIA-1358175),
 NSERC (Grant No SAPPJ-2015-0023), and INFN (Italy).
 A.~F. acknowledges additional support from the TPU (Grant No. LRU-FTI-123-2014)
 and the MSE Program ``Nauka'' (Project No. 3.825.2014/K).
We thank the undergraduate students of Mount Allison University
and The George Washington University for their assistance.
\end{acknowledgments}

\bibliography{literature}

\end{document}